\documentclass[10pt,conference]{IEEEtran}
\usepackage{cite}
\usepackage{amsmath,amssymb,amsfonts}
\usepackage{algorithm}
\usepackage{algpseudocode}
\usepackage{array}
\usepackage{graphicx}
\usepackage{textcomp}
\usepackage{xcolor}
\usepackage{booktabs}
\usepackage{multirow}
\usepackage{stfloats}
\usepackage{tikz}
\usepackage{pgfplots}
\usepackage{pgfplotstable}
\pgfplotsset{compat=1.18}
\usetikzlibrary{arrows.meta,positioning,backgrounds,patterns,decorations.pathreplacing,calc,matrix}
\usepgfplotslibrary{fillbetween}
\usepgfplotslibrary{groupplots}
\usepackage[hyphens]{url}
\usepackage{fancyhdr}
\usepackage{enumitem}
\usepackage{caption}
\usepackage{float}
\usepackage{placeins}

\usepackage{tabularx}
\usepackage{listings}
\usepackage{pdfpages}
\usepackage{newunicodechar}
\usepackage{hyperref}

\usetikzlibrary{shapes.geometric, positioning, arrows.meta}
\tikzset{
  cfblock/.style={draw, rectangle, minimum width=2.5cm, minimum height=0.6cm,
                  align=center, font=\normalsize, inner sep=2pt},
  cfdec/.style={draw, diamond, aspect=2.0, align=center, font=\normalsize,
                inner sep=1pt, minimum width=1.9cm, minimum height=0.6cm},
  cfline/.style={-{Stealth[length=2.4mm]}, thick},
  cflab/.style={font=\normalsize, inner sep=1.5pt}
}

\newunicodechar{▁}{\textunderscore}
\newunicodechar{≤}{$\leq$}
\newunicodechar{→}{$\rightarrow$}

\definecolor{palA}{HTML}{08306B}
\definecolor{palB}{HTML}{2171B5}
\definecolor{palC}{HTML}{6BAED6}
\definecolor{palD}{HTML}{BDD7E7}
\definecolor{palE}{HTML}{EFF3FF}
\definecolor{accentR}{HTML}{A11D24}
\definecolor{accentG}{HTML}{1F6B3A}
\renewcommand{\arraystretch}{1.1}
\definecolor{codebg}{RGB}{248,248,246}
\definecolor{codekw}{RGB}{20,60,140}
\definecolor{codecm}{RGB}{110,110,110}
\lstdefinestyle{paperlst}{%
 backgroundcolor=\color{codebg},basicstyle=\ttfamily\scriptsize,
 keywordstyle=\color{codekw}\bfseries,commentstyle=\color{codecm}\itshape,
 frame=single,framerule=0.3pt,framesep=3pt,breaklines=true,
 columns=fullflexible,keepspaces=true,xleftmargin=3pt,xrightmargin=2pt,
 aboveskip=3pt,belowskip=2pt}
\definecolor{cblue}{HTML}{1C3A6E}
\definecolor{cred}{HTML}{8A1C1C}
\definecolor{cgreen}{HTML}{2E7D46}
\definecolor{clblue}{HTML}{9DB3D4}
\definecolor{ctan}{HTML}{C9D4B7}
\definecolor{corange}{HTML}{B85C00}
\definecolor{cpurple}{HTML}{5B3A8A}

\newcommand{\sys}{\textsc{CacheFit}}
\newcommand{\shippp}{\textsc{SHiP++}}
\newcommand{\hawkeye}{\textsc{Hawkeye}}
\newcommand{\mpppb}{\textsc{MPPPB}}
\newcommand{\champsim}{\textsc{ChampSim}}
\newcommand{\ps}{\,ps}

\newcommand{\sip}{\mbox{SHiP++}}

\newcommand{\harcom}{\textsc{Harcom}}
\newcommand{\code}[1]{\texttt{\small #1}}

\newcommand{\ins}[2]{\par\smallskip\noindent\textbf{Insight~#1:}~\emph{#2}\ }

\title{\LARGE{\textsc{CacheFit}: The Rising Cost of Cache Replacement and
Constraint-Driven Redesign of Last-Level Cache Replacement Policies}}

\author{
  \IEEEauthorblockN{Bita Aslrousta, Kaushal Mhapsekar, Darsh Asher, \\ Azam Ghanbari, Joshua Kalyanapu, and Samira Mirbagher Ajorpaz}
 \IEEEauthorblockA{
    Department of Electrical and Computer Engineering \\
    North Carolina State University, Raleigh, NC, USA \\
    \{baslrou, kmhapse, dkasher, aghanba2, jkalyan, smirbag\}@ncsu.edu
  }
}

\begin{document}
\maketitle
\thispagestyle{plain}
\pagestyle{plain}

\begin{abstract}

The latency, power, and area costs of LLC replacement remain poorly explored not
because simulators cannot be instrumented---event counters feeding a power
model are routine---but because no cost model takes a replacement policy as
input. CACTI and McPAT price regular structures, a cache or a core, not the
small, oddly shaped arrays and dependent logic a policy is made of, and no
event count reveals a victim-selection path's logic depth. Without a model,
cost has never been an objective a policy search could optimize.

In practice, a policy is first selected in simulation for its IPC, and its
cost is reduced later in RTL, after the algorithm is fixed. Cost thus enters
the design too late to shape it. This paper asks what happens when cost is
present from the start.

We build \sys{}, the first policy-design loop for LLC replacement with circuit-level cost inside the search objective. \sys{} connects
ChampSim to \textsc{HARCOM}, a circuit-level cost model, and to a large language model that
proposes policy variants. Every candidate is scored on latency, area, power,
and speedup while its algorithm is still being written. The designer sets the area, power, and latency
budgets, and every proposal and every lesson the search learns is stored as
readable text that the designer can check and edit.

We apply \sys{} to two published policies. From MPPPB, the most expensive
baseline in power and transistors, it produces Fit-MPPPB: $60.3\%$ less
storage, $56.3\%$ less static power, and $58.9\%$ fewer transistors, with
$1.1\%$ higher IPC and a hit rate $7.5$ percentage points higher on SPEC
CPU2017. From SHiP++, the highest-IPC baseline, it produces Fit-SHiP++:
$56\%$ less storage, $61\%$ less static power, and $56\%$ fewer transistors,
for $1.54\%$ lower IPC and a hit rate $2.1$ points lower. A policy designed
with its cost in view can therefore be far cheaper to build at similar or
better performance.

\end{abstract}

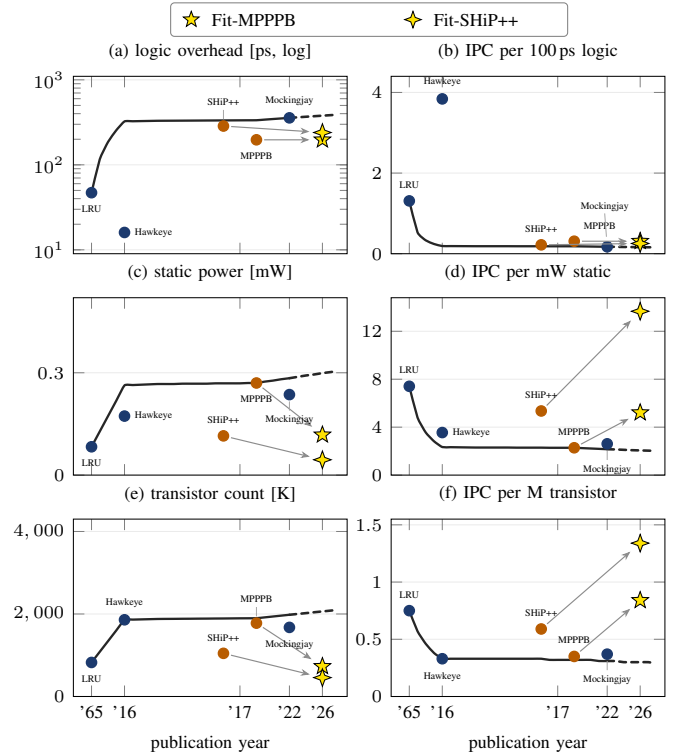
\begin{figure}[!t]
\centering
\resizebox{\columnwidth}{!}{%
\begin{tikzpicture}[font=\small]
\begin{groupplot}[
  group style={group size=2 by 3, horizontal sep=6mm, vertical sep=6mm, x descriptions at=edge bottom},
  width=0.60\columnwidth, height=4.0cm,
  xmin=0.45, xmax=8.75,
  xtick={1,2,5.5,7,8}, xticklabels={'65,'16,'17,'22,'26},
  tick label style={font=\scriptsize}, label style={font=\scriptsize},
  title style={font=\scriptsize, yshift=-3pt},
  ymajorgrids, grid style={black!8},
  clip=false, clip mode=individual,
  every axis plot/.append style={line width=0.5pt},
]
\nextgroupplot[title={(a) logic overhead [ps, log]},
  ymode=log, ymin=9, ymax=1100, ytick={10,100,1000},
  yticklabels={$10^1$,$10^2$,$10^3$}]
\addplot[black!85, line width=0.9pt, line join=round, line cap=round, smooth, tension=0.3, forget plot] coordinates {(1.000,47) (1.250,116.3) (1.500,185.7) (1.750,255) (2.000,324.4) (2.250,325.7) (2.500,327.1) (2.750,328.4) (3.000,329.8) (3.250,330.2) (3.500,330.6) (3.750,331) (4.000,331.4) (4.250,331.8) (4.500,332.3) (4.750,332.7) (5.000,333.1) (5.250,333.5) (5.500,333.9) (5.750,334.3) (6.000,334.7) (6.250,340.3) (6.500,345.9) (6.750,351.4) (7.000,357)};
\addplot[black!85, line width=1pt, line join=round, line cap=round, smooth, tension=0.3, dash pattern=on 2.6pt off 2.1pt, forget plot] coordinates {(7.000,357) (7.250,362.4) (7.500,367.9) (7.750,373.3) (8.000,378.8) (8.250,384.2) (8.450,388.5)};
\addplot[only marks, mark=*, mark size=2pt, cblue, forget plot] coordinates {(1,47)}; %
\addplot[only marks, mark=*, mark size=2pt, cblue, forget plot] coordinates {(2,16)}; %
\addplot[only marks, mark=*, mark size=2pt, corange, forget plot] coordinates {(5,286)}; %
\addplot[only marks, mark=*, mark size=2pt, corange, forget plot] coordinates {(6,197)}; %
\addplot[only marks, mark=*, mark size=2pt, cblue, forget plot] coordinates {(7,357)}; %
\draw[fill=yellow!85!orange, draw=black, line width=0.4pt] ([shift={(0.00pt,3.40pt)}]axis cs:8,197) -- ([shift={(-0.84pt,1.16pt)}]axis cs:8,197) -- ([shift={(-3.23pt,1.05pt)}]axis cs:8,197) -- ([shift={(-1.36pt,-0.44pt)}]axis cs:8,197) -- ([shift={(-2.00pt,-2.75pt)}]axis cs:8,197) -- ([shift={(-0.00pt,-1.43pt)}]axis cs:8,197) -- ([shift={(2.00pt,-2.75pt)}]axis cs:8,197) -- ([shift={(1.36pt,-0.44pt)}]axis cs:8,197) -- ([shift={(3.23pt,1.05pt)}]axis cs:8,197) -- ([shift={(0.84pt,1.16pt)}]axis cs:8,197) -- cycle; %
\draw[fill=yellow!85!orange, draw=black, line width=0.4pt] ([shift={(0.00pt,3.40pt)}]axis cs:8,240) -- ([shift={(-0.91pt,0.91pt)}]axis cs:8,240) -- ([shift={(-3.40pt,0.00pt)}]axis cs:8,240) -- ([shift={(-0.91pt,-0.91pt)}]axis cs:8,240) -- ([shift={(-0.00pt,-3.40pt)}]axis cs:8,240) -- ([shift={(0.91pt,-0.91pt)}]axis cs:8,240) -- ([shift={(3.40pt,-0.00pt)}]axis cs:8,240) -- ([shift={(0.91pt,0.91pt)}]axis cs:8,240) -- cycle; %
\draw[-{Stealth[length=3.6pt]}, black!45, line width=0.45pt, shorten <=3pt, shorten >=5pt] (axis cs:6,197) -- (axis cs:8,197);
\draw[-{Stealth[length=3.6pt]}, black!45, line width=0.45pt, shorten <=3pt, shorten >=5pt] (axis cs:5,286) -- (axis cs:8,240);
\node[text=black, font=\scriptsize, scale=0.55, anchor=north] at (axis cs:1.000,38.19) {LRU};
\node[text=black, font=\scriptsize, scale=0.55, anchor=west] at (axis cs:2.166,16) {Hawkeye};
\node[text=black, font=\scriptsize, scale=0.55, anchor=south] at (axis cs:5.000,442.7) {\sip{}};
\draw[black!35, line width=0.3pt] (axis cs:5.000,442.7) -- (axis cs:5.000,343.6);
\node[text=black, font=\scriptsize, scale=0.55, anchor=north] at (axis cs:6.000,160.1) {MPPPB};
\node[text=black, font=\scriptsize, scale=0.55, anchor=south] at (axis cs:7.000,439.3) {Mockingjay};
\nextgroupplot[title={(b) IPC per 100\,ps logic},
  ymin=0, ymax=4.4, ytick={0,2,4}]
\addplot[black!85, line width=0.9pt, line join=round, line cap=round, smooth, tension=0.3, forget plot] coordinates {(1.000,1.309) (1.250,0.529) (1.500,0.331) (1.750,0.241) (2.000,0.190) (2.250,0.189) (2.500,0.188) (2.750,0.187) (3.000,0.186) (3.250,0.186) (3.500,0.186) (3.750,0.186) (4.000,0.186) (4.250,0.185) (4.500,0.185) (4.750,0.185) (5.000,0.185) (5.250,0.184) (5.500,0.184) (5.750,0.184) (6.000,0.184) (6.250,0.181) (6.500,0.178) (6.750,0.175) (7.000,0.172)};
\addplot[black!85, line width=1pt, line join=round, line cap=round, smooth, tension=0.3, dash pattern=on 2.6pt off 2.1pt, forget plot] coordinates {(7.000,0.172) (7.250,0.170) (7.500,0.167) (7.750,0.165) (8.000,0.162) (8.250,0.160) (8.450,0.158)};
\addplot[only marks, mark=*, mark size=2pt, cblue, forget plot] coordinates {(1,1.310)}; %
\addplot[only marks, mark=*, mark size=2pt, cblue, forget plot] coordinates {(2,3.840)}; %
\addplot[only marks, mark=*, mark size=2pt, corange, forget plot] coordinates {(5,0.220)}; %
\addplot[only marks, mark=*, mark size=2pt, corange, forget plot] coordinates {(6,0.310)}; %
\addplot[only marks, mark=*, mark size=2pt, cblue, forget plot] coordinates {(7,0.172)}; %
\draw[fill=yellow!85!orange, draw=black, line width=0.4pt] ([shift={(0.00pt,3.40pt)}]axis cs:8,0.310) -- ([shift={(-0.84pt,1.16pt)}]axis cs:8,0.310) -- ([shift={(-3.23pt,1.05pt)}]axis cs:8,0.310) -- ([shift={(-1.36pt,-0.44pt)}]axis cs:8,0.310) -- ([shift={(-2.00pt,-2.75pt)}]axis cs:8,0.310) -- ([shift={(-0.00pt,-1.43pt)}]axis cs:8,0.310) -- ([shift={(2.00pt,-2.75pt)}]axis cs:8,0.310) -- ([shift={(1.36pt,-0.44pt)}]axis cs:8,0.310) -- ([shift={(3.23pt,1.05pt)}]axis cs:8,0.310) -- ([shift={(0.84pt,1.16pt)}]axis cs:8,0.310) -- cycle; %
\draw[fill=yellow!85!orange, draw=black, line width=0.4pt] ([shift={(0.00pt,3.40pt)}]axis cs:8,0.254) -- ([shift={(-0.91pt,0.91pt)}]axis cs:8,0.254) -- ([shift={(-3.40pt,0.00pt)}]axis cs:8,0.254) -- ([shift={(-0.91pt,-0.91pt)}]axis cs:8,0.254) -- ([shift={(-0.00pt,-3.40pt)}]axis cs:8,0.254) -- ([shift={(0.91pt,-0.91pt)}]axis cs:8,0.254) -- ([shift={(3.40pt,-0.00pt)}]axis cs:8,0.254) -- ([shift={(0.91pt,0.91pt)}]axis cs:8,0.254) -- cycle; %
\draw[-{Stealth[length=3.6pt]}, black!45, line width=0.45pt, shorten <=3pt, shorten >=5pt] (axis cs:6,0.310) -- (axis cs:8,0.310);
\draw[-{Stealth[length=3.6pt]}, black!45, line width=0.45pt, shorten <=3pt, shorten >=5pt] (axis cs:5,0.220) -- (axis cs:8,0.254);
\node[text=black, font=\scriptsize, scale=0.55, anchor=south] at (axis cs:1.000,1.500) {LRU};
\node[text=black, font=\scriptsize, scale=0.55, anchor=south] at (axis cs:2.000,4.030) {Hawkeye};
\node[text=black, font=\scriptsize, scale=0.55, anchor=south] at (axis cs:5.000,0.410) {\sip{}};
\node[text=black, font=\scriptsize, scale=0.55, anchor=south west] at (axis cs:6.138,0.490) {MPPPB};
\draw[black!35, line width=0.3pt] (axis cs:6.924,1.000) -- (axis cs:6.977,0.420);
\node[text=black, font=\scriptsize, scale=0.55, anchor=center] at (axis cs:6.924,1.165) {Mockingjay};
\nextgroupplot[title={(c) static power [mW]},
  ymin=0, ymax=0.52, ytick={0,0.3}]
\addplot[black!85, line width=0.9pt, line join=round, line cap=round, smooth, tension=0.3, forget plot] coordinates {(1.000,0.083) (1.250,0.128) (1.500,0.173) (1.750,0.218) (2.000,0.263) (2.250,0.264) (2.500,0.265) (2.750,0.266) (3.000,0.267) (3.250,0.267) (3.500,0.267) (3.750,0.268) (4.000,0.268) (4.250,0.268) (4.500,0.268) (4.750,0.269) (5.000,0.269) (5.250,0.269) (5.500,0.269) (5.750,0.270) (6.000,0.270) (6.250,0.274) (6.500,0.277) (6.750,0.281) (7.000,0.284)};
\addplot[black!85, line width=1pt, line join=round, line cap=round, smooth, tension=0.3, dash pattern=on 2.6pt off 2.1pt, forget plot] coordinates {(7.000,0.284) (7.250,0.288) (7.500,0.292) (7.750,0.295) (8.000,0.299) (8.250,0.302) (8.450,0.305)};
\addplot[only marks, mark=*, mark size=2pt, cblue, forget plot] coordinates {(1,0.083)}; %
\addplot[only marks, mark=*, mark size=2pt, cblue, forget plot] coordinates {(2,0.173)}; %
\addplot[only marks, mark=*, mark size=2pt, corange, forget plot] coordinates {(5,0.115)}; %
\addplot[only marks, mark=*, mark size=2pt, corange, forget plot] coordinates {(6,0.270)}; %
\addplot[only marks, mark=*, mark size=2pt, cblue, forget plot] coordinates {(7,0.236)}; %
\draw[fill=yellow!85!orange, draw=black, line width=0.4pt] ([shift={(0.00pt,3.40pt)}]axis cs:8,0.118) -- ([shift={(-0.84pt,1.16pt)}]axis cs:8,0.118) -- ([shift={(-3.23pt,1.05pt)}]axis cs:8,0.118) -- ([shift={(-1.36pt,-0.44pt)}]axis cs:8,0.118) -- ([shift={(-2.00pt,-2.75pt)}]axis cs:8,0.118) -- ([shift={(-0.00pt,-1.43pt)}]axis cs:8,0.118) -- ([shift={(2.00pt,-2.75pt)}]axis cs:8,0.118) -- ([shift={(1.36pt,-0.44pt)}]axis cs:8,0.118) -- ([shift={(3.23pt,1.05pt)}]axis cs:8,0.118) -- ([shift={(0.84pt,1.16pt)}]axis cs:8,0.118) -- cycle; %
\draw[fill=yellow!85!orange, draw=black, line width=0.4pt] ([shift={(0.00pt,3.40pt)}]axis cs:8,0.045) -- ([shift={(-0.91pt,0.91pt)}]axis cs:8,0.045) -- ([shift={(-3.40pt,0.00pt)}]axis cs:8,0.045) -- ([shift={(-0.91pt,-0.91pt)}]axis cs:8,0.045) -- ([shift={(-0.00pt,-3.40pt)}]axis cs:8,0.045) -- ([shift={(0.91pt,-0.91pt)}]axis cs:8,0.045) -- ([shift={(3.40pt,-0.00pt)}]axis cs:8,0.045) -- ([shift={(0.91pt,0.91pt)}]axis cs:8,0.045) -- cycle; %
\draw[-{Stealth[length=3.6pt]}, black!45, line width=0.45pt, shorten <=3pt, shorten >=5pt] (axis cs:6,0.270) -- (axis cs:8,0.118);
\draw[-{Stealth[length=3.6pt]}, black!45, line width=0.45pt, shorten <=3pt, shorten >=5pt] (axis cs:5,0.115) -- (axis cs:8,0.045);
\node[text=black, font=\scriptsize, scale=0.55, anchor=north] at (axis cs:1.000,0.061) {LRU};
\node[text=black, font=\scriptsize, scale=0.55, anchor=west] at (axis cs:2.166,0.173) {Hawkeye};
\node[text=black, font=\scriptsize, scale=0.55, anchor=south] at (axis cs:5.000,0.137) {\sip{}};
\node[text=black, font=\scriptsize, scale=0.55, anchor=north] at (axis cs:6.000,0.248) {MPPPB};
\node[text=black, font=\scriptsize, scale=0.55, anchor=north] at (axis cs:7.000,0.189) {Mockingjay};
\draw[black!35, line width=0.3pt] (axis cs:7.000,0.189) -- (axis cs:7.000,0.216);
\nextgroupplot[title={(d) IPC per mW static},
  ymin=0, ymax=14.8, ytick={0,4,8,12}]
\addplot[black!85, line width=0.9pt, line join=round, line cap=round, smooth, tension=0.3, forget plot] coordinates {(1.000,7.41) (1.250,4.80) (1.500,3.55) (1.750,2.82) (2.000,2.34) (2.250,2.33) (2.500,2.32) (2.750,2.31) (3.000,2.30) (3.250,2.30) (3.500,2.30) (3.750,2.30) (4.000,2.30) (4.250,2.29) (4.500,2.29) (4.750,2.29) (5.000,2.29) (5.250,2.28) (5.500,2.28) (5.750,2.28) (6.000,2.28) (6.250,2.25) (6.500,2.22) (6.750,2.19) (7.000,2.16)};
\addplot[black!85, line width=1pt, line join=round, line cap=round, smooth, tension=0.3, dash pattern=on 2.6pt off 2.1pt, forget plot] coordinates {(7.000,2.16) (7.250,2.14) (7.500,2.11) (7.750,2.08) (8.000,2.06) (8.250,2.04) (8.450,2.02)};
\addplot[only marks, mark=*, mark size=2pt, cblue, forget plot] coordinates {(1,7.41)}; %
\addplot[only marks, mark=*, mark size=2pt, cblue, forget plot] coordinates {(2,3.55)}; %
\addplot[only marks, mark=*, mark size=2pt, corange, forget plot] coordinates {(5,5.35)}; %
\addplot[only marks, mark=*, mark size=2pt, corange, forget plot] coordinates {(6,2.28)}; %
\addplot[only marks, mark=*, mark size=2pt, cblue, forget plot] coordinates {(7,2.61)}; %
\draw[fill=yellow!85!orange, draw=black, line width=0.4pt] ([shift={(0.00pt,3.40pt)}]axis cs:8,5.21) -- ([shift={(-0.84pt,1.16pt)}]axis cs:8,5.21) -- ([shift={(-3.23pt,1.05pt)}]axis cs:8,5.21) -- ([shift={(-1.36pt,-0.44pt)}]axis cs:8,5.21) -- ([shift={(-2.00pt,-2.75pt)}]axis cs:8,5.21) -- ([shift={(-0.00pt,-1.43pt)}]axis cs:8,5.21) -- ([shift={(2.00pt,-2.75pt)}]axis cs:8,5.21) -- ([shift={(1.36pt,-0.44pt)}]axis cs:8,5.21) -- ([shift={(3.23pt,1.05pt)}]axis cs:8,5.21) -- ([shift={(0.84pt,1.16pt)}]axis cs:8,5.21) -- cycle; %
\draw[fill=yellow!85!orange, draw=black, line width=0.4pt] ([shift={(0.00pt,3.40pt)}]axis cs:8,13.67) -- ([shift={(-0.91pt,0.91pt)}]axis cs:8,13.67) -- ([shift={(-3.40pt,0.00pt)}]axis cs:8,13.67) -- ([shift={(-0.91pt,-0.91pt)}]axis cs:8,13.67) -- ([shift={(-0.00pt,-3.40pt)}]axis cs:8,13.67) -- ([shift={(0.91pt,-0.91pt)}]axis cs:8,13.67) -- ([shift={(3.40pt,-0.00pt)}]axis cs:8,13.67) -- ([shift={(0.91pt,0.91pt)}]axis cs:8,13.67) -- cycle; %
\draw[-{Stealth[length=3.6pt]}, black!45, line width=0.45pt, shorten <=3pt, shorten >=5pt] (axis cs:6,2.28) -- (axis cs:8,5.21);
\draw[-{Stealth[length=3.6pt]}, black!45, line width=0.45pt, shorten <=3pt, shorten >=5pt] (axis cs:5,5.35) -- (axis cs:8,13.67);
\node[text=black, font=\scriptsize, scale=0.55, anchor=south] at (axis cs:1.000,8.05) {LRU};
\node[text=black, font=\scriptsize, scale=0.55, anchor=west] at (axis cs:2.166,3.55) {Hawkeye};
\node[text=black, font=\scriptsize, scale=0.55, anchor=south] at (axis cs:5.000,5.99) {\sip{}};
\node[text=black, font=\scriptsize, scale=0.55, anchor=south] at (axis cs:6.000,2.92) {MPPPB};
\node[text=black, font=\scriptsize, scale=0.55, anchor=north] at (axis cs:7.000,1.26) {Mockingjay};
\draw[black!35, line width=0.3pt] (axis cs:7.000,1.26) -- (axis cs:7.000,2.04);
\nextgroupplot[title={(e) transistor count [K]},
  xlabel={publication year},
  ymin=0, ymax=4300, ytick={0,2000,4000},
  yticklabel style={/pgf/number format/.cd, fixed, 1000 sep={,}}]
\addplot[black!85, line width=0.9pt, line join=round, line cap=round, smooth, tension=0.3, forget plot] coordinates {(1.000,825) (1.250,1084) (1.500,1342) (1.750,1601) (2.000,1859) (2.250,1864) (2.500,1869) (2.750,1874) (3.000,1879) (3.250,1881) (3.500,1882) (3.750,1884) (4.000,1885) (4.250,1887) (4.500,1888) (4.750,1890) (5.000,1891) (5.250,1893) (5.500,1895) (5.750,1896) (6.000,1898) (6.250,1918) (6.500,1939) (6.750,1960) (7.000,1981)};
\addplot[black!85, line width=1pt, line join=round, line cap=round, smooth, tension=0.3, dash pattern=on 2.6pt off 2.1pt, forget plot] coordinates {(7.000,1981) (7.250,2001) (7.500,2021) (7.750,2042) (8.000,2062) (8.250,2082) (8.450,2098)};
\addplot[only marks, mark=*, mark size=2pt, cblue, forget plot] coordinates {(1,825)}; %
\addplot[only marks, mark=*, mark size=2pt, cblue, forget plot] coordinates {(2,1859)}; %
\addplot[only marks, mark=*, mark size=2pt, corange, forget plot] coordinates {(5,1046)}; %
\addplot[only marks, mark=*, mark size=2pt, corange, forget plot] coordinates {(6,1778)}; %
\addplot[only marks, mark=*, mark size=2pt, cblue, forget plot] coordinates {(7,1675)}; %
\draw[fill=yellow!85!orange, draw=black, line width=0.4pt] ([shift={(0.00pt,3.40pt)}]axis cs:8,730) -- ([shift={(-0.84pt,1.16pt)}]axis cs:8,730) -- ([shift={(-3.23pt,1.05pt)}]axis cs:8,730) -- ([shift={(-1.36pt,-0.44pt)}]axis cs:8,730) -- ([shift={(-2.00pt,-2.75pt)}]axis cs:8,730) -- ([shift={(-0.00pt,-1.43pt)}]axis cs:8,730) -- ([shift={(2.00pt,-2.75pt)}]axis cs:8,730) -- ([shift={(1.36pt,-0.44pt)}]axis cs:8,730) -- ([shift={(3.23pt,1.05pt)}]axis cs:8,730) -- ([shift={(0.84pt,1.16pt)}]axis cs:8,730) -- cycle; %
\draw[fill=yellow!85!orange, draw=black, line width=0.4pt] ([shift={(0.00pt,3.40pt)}]axis cs:8,458) -- ([shift={(-0.91pt,0.91pt)}]axis cs:8,458) -- ([shift={(-3.40pt,0.00pt)}]axis cs:8,458) -- ([shift={(-0.91pt,-0.91pt)}]axis cs:8,458) -- ([shift={(-0.00pt,-3.40pt)}]axis cs:8,458) -- ([shift={(0.91pt,-0.91pt)}]axis cs:8,458) -- ([shift={(3.40pt,-0.00pt)}]axis cs:8,458) -- ([shift={(0.91pt,0.91pt)}]axis cs:8,458) -- cycle; %
\draw[-{Stealth[length=3.6pt]}, black!45, line width=0.45pt, shorten <=3pt, shorten >=5pt] (axis cs:6,1778) -- (axis cs:8,730);
\draw[-{Stealth[length=3.6pt]}, black!45, line width=0.45pt, shorten <=3pt, shorten >=5pt] (axis cs:5,1046) -- (axis cs:8,458);
\node[text=black, font=\scriptsize, scale=0.55, anchor=north] at (axis cs:1.000,639) {LRU};
\node[text=black, font=\scriptsize, scale=0.55, anchor=south] at (axis cs:2.000,2045) {Hawkeye};
\node[text=black, font=\scriptsize, scale=0.55, anchor=south] at (axis cs:5.000,1232) {\sip{}};
\node[text=black, font=\scriptsize, scale=0.55, anchor=south] at (axis cs:6.000,2169) {MPPPB};
\draw[black!35, line width=0.3pt] (axis cs:6.000,2169) -- (axis cs:6.000,1942);
\node[text=black, font=\scriptsize, scale=0.55, anchor=north] at (axis cs:7.000,1489) {Mockingjay};
\nextgroupplot[title={(f) IPC per M transistor},
  xlabel={publication year},
  ymin=0, ymax=1.55, ytick={0,0.5,1,1.5}]
\addplot[black!85, line width=0.9pt, line join=round, line cap=round, smooth, tension=0.3, forget plot] coordinates {(1.000,0.75) (1.250,0.57) (1.500,0.46) (1.750,0.38) (2.000,0.33) (2.250,0.33) (2.500,0.33) (2.750,0.33) (3.000,0.33) (3.250,0.33) (3.500,0.33) (3.750,0.33) (4.000,0.33) (4.250,0.33) (4.500,0.33) (4.750,0.33) (5.000,0.33) (5.250,0.32) (5.500,0.32) (5.750,0.32) (6.000,0.32) (6.250,0.32) (6.500,0.32) (6.750,0.31) (7.000,0.31)};
\addplot[black!85, line width=1pt, line join=round, line cap=round, smooth, tension=0.3, dash pattern=on 2.6pt off 2.1pt, forget plot] coordinates {(7.000,0.31) (7.250,0.31) (7.500,0.30) (7.750,0.30) (8.000,0.30) (8.250,0.30) (8.450,0.29)};
\addplot[only marks, mark=*, mark size=2pt, cblue, forget plot] coordinates {(1,0.75)}; %
\addplot[only marks, mark=*, mark size=2pt, cblue, forget plot] coordinates {(2,0.33)}; %
\addplot[only marks, mark=*, mark size=2pt, corange, forget plot] coordinates {(5,0.59)}; %
\addplot[only marks, mark=*, mark size=2pt, corange, forget plot] coordinates {(6,0.35)}; %
\addplot[only marks, mark=*, mark size=2pt, cblue, forget plot] coordinates {(7,0.37)}; %
\draw[fill=yellow!85!orange, draw=black, line width=0.4pt] ([shift={(0.00pt,3.40pt)}]axis cs:8,0.84) -- ([shift={(-0.84pt,1.16pt)}]axis cs:8,0.84) -- ([shift={(-3.23pt,1.05pt)}]axis cs:8,0.84) -- ([shift={(-1.36pt,-0.44pt)}]axis cs:8,0.84) -- ([shift={(-2.00pt,-2.75pt)}]axis cs:8,0.84) -- ([shift={(-0.00pt,-1.43pt)}]axis cs:8,0.84) -- ([shift={(2.00pt,-2.75pt)}]axis cs:8,0.84) -- ([shift={(1.36pt,-0.44pt)}]axis cs:8,0.84) -- ([shift={(3.23pt,1.05pt)}]axis cs:8,0.84) -- ([shift={(0.84pt,1.16pt)}]axis cs:8,0.84) -- cycle; %
\draw[fill=yellow!85!orange, draw=black, line width=0.4pt] ([shift={(0.00pt,3.40pt)}]axis cs:8,1.34) -- ([shift={(-0.91pt,0.91pt)}]axis cs:8,1.34) -- ([shift={(-3.40pt,0.00pt)}]axis cs:8,1.34) -- ([shift={(-0.91pt,-0.91pt)}]axis cs:8,1.34) -- ([shift={(-0.00pt,-3.40pt)}]axis cs:8,1.34) -- ([shift={(0.91pt,-0.91pt)}]axis cs:8,1.34) -- ([shift={(3.40pt,-0.00pt)}]axis cs:8,1.34) -- ([shift={(0.91pt,0.91pt)}]axis cs:8,1.34) -- cycle; %
\draw[-{Stealth[length=3.6pt]}, black!45, line width=0.45pt, shorten <=3pt, shorten >=5pt] (axis cs:6,0.35) -- (axis cs:8,0.84);
\draw[-{Stealth[length=3.6pt]}, black!45, line width=0.45pt, shorten <=3pt, shorten >=5pt] (axis cs:5,0.59) -- (axis cs:8,1.34);
\node[text=black, font=\scriptsize, scale=0.55, anchor=south] at (axis cs:1.000,0.82) {LRU};
\node[text=black, font=\scriptsize, scale=0.55, anchor=north] at (axis cs:2.000,0.26) {Hawkeye};
\node[text=black, font=\scriptsize, scale=0.55, anchor=south] at (axis cs:5.000,0.66) {\sip{}};
\node[text=black, font=\scriptsize, scale=0.55, anchor=south] at (axis cs:6.000,0.42) {MPPPB};
\node[text=black, font=\scriptsize, scale=0.55, anchor=north] at (axis cs:7.000,0.23) {Mockingjay};
\draw[black!35, line width=0.3pt] (axis cs:7.000,0.23) -- (axis cs:7.000,0.31);
\end{groupplot}
\coordinate (LG) at ([xshift=47pt, yshift=30pt]$(group c1r1.north)!0.5!(group c2r1.north)$);
\draw[black!50, line width=0.4pt, rounded corners=1.5pt]
  ([shift={(-124pt,-14.5pt)}]LG) rectangle ([shift={(30pt,-1.5pt)}]LG);
\draw[fill=yellow!85!orange, draw=black, line width=0.4pt] ([shift={(-116.00pt,-4.80pt)}]LG) -- ([shift={(-116.79pt,-6.91pt)}]LG) -- ([shift={(-119.04pt,-7.01pt)}]LG) -- ([shift={(-117.28pt,-8.42pt)}]LG) -- ([shift={(-117.88pt,-10.59pt)}]LG) -- ([shift={(-116.00pt,-9.34pt)}]LG) -- ([shift={(-114.12pt,-10.59pt)}]LG) -- ([shift={(-114.72pt,-8.42pt)}]LG) -- ([shift={(-112.96pt,-7.01pt)}]LG) -- ([shift={(-115.21pt,-6.91pt)}]LG) -- cycle;
\node[anchor=west, font=\scriptsize] at ([shift={(-111.5pt,-8.0pt)}]LG) {Fit-MPPPB};
\draw[fill=yellow!85!orange, draw=black, line width=0.4pt] ([shift={(-30.00pt,-4.80pt)}]LG) -- ([shift={(-30.86pt,-7.14pt)}]LG) -- ([shift={(-33.20pt,-8.00pt)}]LG) -- ([shift={(-30.86pt,-8.86pt)}]LG) -- ([shift={(-30.00pt,-11.20pt)}]LG) -- ([shift={(-29.14pt,-8.86pt)}]LG) -- ([shift={(-26.80pt,-8.00pt)}]LG) -- ([shift={(-29.14pt,-7.14pt)}]LG) -- cycle;
\node[anchor=west, font=\scriptsize] at ([shift={(-25.5pt,-8.0pt)}]LG) {Fit-\sip{}};
\end{tikzpicture}
}%
\caption{Hardware cost per policy (left) and IPC returned per unit of that
cost vs.\ publication
year.}
\label{fig:cost-eff-column}
\end{figure}

\section{Introduction}
\label{sec:intro}

The gap between cache hit latency and memory latency continues to widen,
so a replacement policy that removes misses translates directly into
performance. What that policy costs in latency, power, and area is rarely
studied in depth. Replacement is measured in speedup, and simulation is treated as sufficient, on the assumption that cost is a separate axis to be settled
later. It is not separate. Redesigning for cost and feasibility changes what a policy achieves: in neural predictors it can raise IPC~\cite{mpppb}, and in
counter- and table-based predictors it can cut power outright~\cite{rrip,red,ship}. 

This paper measures how individual design decisions inside a replacement
policy---table counts, counter widths, sampler shape, victim-selection
structure---move latency, power, area, and speedup. We release the circuit
model and the search tool that uses it, with worked examples showing how to
cut a policy's cost substantially while holding or improving its speedup. The search is driven by a large language model, but the designer
holds the constraints: every step of the search path is recorded, auditable,
and editable.

The literature typically reports a replacement policy's cost as one number:
kilobytes of metadata~\cite{ship,shippp,sdbp,lime,hawkeye,mpppb}. Latency, when
it appears at all, is estimated by proxy---counting accesses to predictor
tables and treating that as delay~\cite{ajorpaz2022chirp}. Until now there has been no tool that could measure a policy at
the circuit level, so the field has designed around the one axis it could see.
We evaluate six LLC policies with a 5\,nm FinFET circuit model (Table \ref{tab:hw}) and show that
metadata size, the number prior work has relied on, not only misranks hardware
cost but can also hurt IPC when it guides redesign.

Take the two clearest cases. \sip{}~\cite{shippp} declares $20$\,KB of metadata
and Hawkeye~\cite{hawkeye} declares $36$\,KB, so by the field's metric \sip{} is
the cheaper design by a comfortable margin. In silicon the ordering reverses.
After its metadata read, \sip{} runs $286$\,ps of dependent combinational
logic---the deepest chain of any policy we measure, $6.1\times$ LRU's
$47$\,ps---while Hawkeye runs $16$\,ps (a lower bound, \S\ref{sec:finding2}), the shallowest. Static power inverts as
well: a kilobyte of metadata leaks between $0.0048$\,mW in Hawkeye and
$0.0083$\,mW in RED, a $1.7\times$ spread that follows array geometry rather
than capacity. The metric does not merely lose precision here; it points the
opposite way.

MPPPB~\cite{mpppb}, our neural baseline, sits above LRU on storage, static
power, and transistors, and below LRU on each corresponding
efficiency \textbf{(Fig.~\ref{fig:cost-eff-column})}---IPC per kilobyte, per milliwatt, per million transistors. It costs
$2.1$--$3.3\times$ LRU and returns $2.1$--$3.3\times$ less performance per unit
of that cost, while delivering $0.76\%$ \emph{lower} SPEC~CPU2017 IPC.
Hawkeye~\cite{hawkeye} is in the same position: costlier than LRU on all three
axes, less efficient on all three. No individual policy year dominates the
next---
Hawkeye (2016) leaks more than \sip{} (2017)~\cite{shippp}---so we claim no
steady year-on-year climb. But the returns have not merely diminished; on at
least one axis they have gone negative. The field's two most elaborate
mechanisms cost more than LRU on storage, power, and transistors, and one of
them performs worse than LRU while doing it. None of the papers introducing
them reports the axes on which this is visible. Continue designing this way and
a policy in 2030 costs roughly $35$\,KB of metadata and $0.225$\,mW of static
power (Fig.~\ref{fig:cost-eff-column}e,f). The cheapest policy in this paper already sits $75\%$ below that storage figure and $80\%$ below the power figure. 

\begin{table*}[!htbp]
\renewcommand{\arraystretch}{1.18}
\centering
\small
\setlength{\tabcolsep}{3.2pt}
\begin{tabular}{r l r r r r r r r r r r r}
\toprule
Year & Policy & Lat. & SRAM & Logic & Stor. & Stat. & Xtors & Area & IPC & HR & IPC/KB & IPC/mW\\
     &        & (ps) & (ps) & OH (ps)& (KB)  & (mW)  & (K) & (mm$^2$) & '17 & (\%) & {\scriptsize($\times10^{-2}$)} & \\
\midrule
1967  & LRU              & 316 & 269 & 47  & 16.0  & 0.083 & 825 & 0.0042 & 0.6154 & 49.2 & 3.85 & 7.4\\
2016 & Hawkeye$^\dagger$& 284 & 268 & 16  & 36.0  & 0.173 & 1859 & 0.0100 & 0.6154 & 41.9 & 1.71 & 3.6\\
2017 & RED$^\ast$       & 462 & 326 & 136 & 32 & 0.267 & 1697 & 0.0086 & 0.6179 & 42.7 & 1.93 & 2.3\\
2017 & \sip{}           & 509 & 223 & {\textbf{286}} & 20.0 & {\textbf{0.115}} & 1046 & 0.0059 & 0.6181 & 38.9 & 3.09 & 5.4\\
2017 & MPPPB$^\dagger$ & 284 & 87 & 197 & 33.2  & 0.270 & 1778 & 0.0092 & 0.6107 & 29.0 & 1.84 & 2.3\\
2022 & Mockingjay       & 661 & 304 & 357 & 31.9  & 0.236 & 1675 & 0.0119 & 0.6188 & 42.4 & 1.94 & 2.6\\
\midrule
\textbf{2026} & \textbf{Fit-MPPPB}$^\ddagger$
     & 283 & 86 & 197 & {\textbf{13.2}} & {\textbf{0.118}}
     & {\textbf{730}} & {\textbf{0.0038}} & 0.6174 & {\textbf{36.5}} & {\textbf{4.68}} & {\textbf{5.2}}\\
\textbf{2026} & \textbf{Fit-SHiP++}$^\ddagger$
     & 463 & 223 & 240 & {\textbf{8.8}} & {\textbf{0.045}}
     & {\textbf{458}} & {\textbf{0.0025}} & 0.6086 & 36.8 & {\textbf{6.92}} & {\textbf{13.5}}\\
\textbf{2026} & \textbf{Perf-SHiP++}$^\ddagger$
     & 502 & 244 & 258 & 25.19 & 0.133
     & 1308 & 0.00695 & 0.6165 & 39.0 & 2.45 & 4.64\\
\bottomrule
\end{tabular}
\caption{Hardware cost and Champsim output of SOTA and \sys{} policies sorted by publication year. $^\dagger$Logic overhead is a lower bound from a partially elaborated path. $^\ast$Shared multi-core LLC with $4\times$ the set count; not directly comparable to the single-core rows. $^\ddagger$Produced by \sys{} (this work).}
\label{tab:hw}
\end{table*}

Our results show that constraining MPPPB's cost with \sys{} made it \emph{better}: the search pruned features, and what survived predicted more accurately than the full set. Constraining \sip{} made it cheaper at held accuracy---fitting removes waste, but a signature-based policy has less waste to give. We suspect this is structural. A learned predictor's cost sits in table count and counter width, so pruning acts as regularization; a table-based policy's accuracy sits in the state itself, so shrinking it removes information. If that holds, neural predictors across microarchitecture may be over-provisioned in ways their designers have had no way to see.

\medskip\textbf{Our approach.} \sys{} places the language model inside a governed
generate--evaluate--refine loop, following the generator--evaluator split that
drives FunSearch and AlphaEvolve~\cite{romera2024funsearch,novikov2025alphaevolve},
with a circuit-level cost model in the evaluator. Structure-aware prompting
keeps every candidate conformant to the ChampSim interface; static feasibility
checks reject unsafe or hallucinated code before it reaches simulation; storage
and cost budgets bound the search, forcing the model to trade capacity for
performance rather than grow without limit. A cycle-accurate evaluator supplies
the causal feedback that imitation-trained generators lack, penalizing designs
that raise hit rate while lengthening exposed stalls, and the objective is IPC
and IPC per KB/mW rather than raw hit rate. Outcomes accumulate into a
persistent, human-readable record of what worked and why---which the designer
can read, edit, and redirect mid-search. To our knowledge this is the first
design-exploration loop for replacement policies in which the hardware cost
inside the objective is a circuit-level measurement rather than a bit count.

\medskip We make three contributions:
\begin{itemize}

\item We measure a rising cost trend on three axes for LLC replacement
  policies and show that the field's ranking inverts. Ordering policies by
  metadata misranks them once circuit-level overhead is measured. Hawkeye,
  the most expensive by metadata, is the cheapest in hardware, while \sip{},
  second-cheapest by metadata, is second-most expensive. Metadata gets static
  power's ordering nearly right but mis-scales it by $1.7\times$; of the
  three axes we add, only transistor count follows metadata faithfully.

\item We show with concrete examples why conventional redesign of replacement
  policies for efficiency fails, and trace the failure to its root causes
  (\S\ref{sec:finding1}). Building on that analysis, we propose \sys{}, a
  design-exploration loop that puts an LLM inside \textsc{ChampSim} with \harcom{} in
  the objective, searching implementable policy variants under measured
  circuit-level cost caps. We show why conventional AI search tools such as
  FunSearch fail at this task and how six interventions fix them, and we
  show that \sys{}'s auditable, editable memory lets a designer follow the
  search path and correct it.

\item We walk through three end-to-end case studies of \sys{}---input policy,
  final output, and the bit-level component changes through which each gain
  was won---showing exactly how latency, power, and area efficiency improved
  while IPC and hit rate were preserved.
  (i)~\emph{Constraining cost made the neural policy better.} Applied to
  MPPPB, the most expensive baseline on power and transistors, \sys{} yields
  Fit-MPPPB: storage $-60.3\%$, static power $-56.3\%$, transistors
  $-58.9\%$, area $-58.7\%$, at unchanged latency. Hit rate rises on 18 of 20
  SPEC~CPU2017 workloads ($29.0\%\!\to\!36.5\%$) and IPC on 14 of 20
  ($+1.11\%$ aggregate, concentrated in cactuBSSN)---on a policy now cheaper
  than LRU on storage, transistors, and area.
  (ii)~\emph{The same search holds accuracy on a table-based policy.} Applied
  to \sip{}, the highest-IPC baseline, it yields Fit-SHiP++: storage
  $-56.0\%$, static power $-60.9\%$, transistors $-56.2\%$, area $-57.6\%$,
  for a $1.54\%$ IPC drop---a $2.5\times$ gain in IPC per milliwatt and the
  cheapest policy in the study on power and area.
  (iii)~\emph{Released from its cost cap, \sys{} wins on IPC alone.} Given no
  restricted seed, it evolves Perf-SHiP++, which reaches a harmonic-mean IPC
  of $0.4958$ across all 27 SPEC~CPU2006 workloads, $+0.39\%$ over
  \sip{}---the best of the six baselines.

\end{itemize}

\section{Background \& Related Works}
\label{sec:background}

Growing working sets~\cite{ferdman2012clearing,li2015sparkbench}, the
processor--memory gap~\cite{wulf1995memorywall}, and diminishing returns from
technology scaling~\cite{esmaeilzadeh2011darksilicon} have made on-chip cache
capacity and its management a first-order design concern: caches occupy up to 40\% of die area
and 20\% of chip energy~\cite{won2015cake}. Replacement policy work spans
insertion~\cite{fedorov2013ari,rrip,jimenez2013insertion,khan2014readwrite,dip},
dead-block and reuse
prediction~\cite{sdbp,lai2001deadblock,seshadri2012eaf,teran2016mdpp,ship},
and prefetching~\cite{baer1991preloading,ishii2011access,kadjo2014bfetch,kim2016path,michaud2016bestoffset,pugsley2014sandbox,somogyi2006spatial,srinath2007feedback}.
Across all of it, cost is reported as kilobytes of metadata---not for lack of
interest, but for lack of a tool. ChampSim counts cycles and models no gates.
Circuit-level tools take regular structures, a cache configuration or a core,
not a small metadata array feeding a reduce tree. RTL synthesis would answer
the question, but only once the design is frozen and the answer can no longer
steer it.

\medskip\subsection{Three replacement policy families}
\label{sec:bg-families}
 (i) Recency alone is a sufficient signal. LRU keeps a 4-bit recency counter per
way; our reference point, not a shipping design. RRIP~\cite{rrip} replaces that counter with a 2-bit re-reference
value and evicts the first way at maximum RRPV---the last family confirmed in
silicon~\cite{ivbrevengr,cachequery,reloadrefresh}.

 (ii) The fetching instruction predicts reuse. SHiP~\cite{ship}
hashes the fill PC into a signature that indexes a Signature History Counter
Table (SHCT). \sip{}~\cite{shippp} adds eviction feedback, correcting the
departing block's counter rather than only the hit path.
LIME~\cite{lime} reads the same PC signal but spends it on a
parallel bypass test instead of an insertion depth. RED~\cite{red}
makes that bypass decision for a shared multi-core LLC; its $4\times$ set
count makes it incomparable to the single-core rows, and we flag it wherever
it appears.

 (iii) The eviction target is learned
directly. \hawkeye{}~\cite{hawkeye} reconstructs Belady's decisions offline through
OPTgen, an occupancy-vector sampler, and trains a PC-indexed predictor on the
resulting labels---all of it off the demand path. \mpppb{}~\cite{mpppb}, our
search target, applies the perceptron principle~\cite{jimenez01,perceptron}:
sixteen $256\times6$-bit tables, each hashing a different feature (PC, path
history, address and offset, insertion state, last miss), are read in
parallel, summed, and thresholded three ways for placement, promotion, and
bypass, with sampled training to bound update traffic.

\medskip\subsection{\harcom{}: an analytical circuit model, not a synthesis flow}
\label{sec:bg-harcom}

Every hardware number in this paper comes from \harcom{}~\cite{harcom}, a
closed-form, compile-time-evaluated C++20 circuit model. It composes typed
blocks---\code{ram} arrays, arithmetic and selection primitives---and returns
delay, static power, dynamic energy, transistor count, and capacitance;
sequential blocks sum delay, parallel blocks take the maximum. Every policy is
modeled at the same 5\,nm FinFET target (2\,MB, $2048\times16$-way) with the
same cell library, so cost columns compare without normalization. One SRAM bank's read latency is
\begin{equation}
t_\text{read} = \max\bigl(t_\text{WLSEL}+t_\text{WLRC}+t_\text{BLDLY}+t_\text{SA},\;
                   t_{\text{CMUX}_0}\bigr) + t_{\text{CMUX}_1}
\label{eq:sram}
\end{equation}

where $t_\text{WLSEL}$ (row decode) punishes tall arrays, $t_\text{WLRC}$
(wordline RC) punishes wide ones, $t_\text{BLDLY}$ grows with rows per
bitline, and $t_\text{SA}$ is the sense amplifier. Bank shape is not searched:
\harcom{} computes a roughly square row/column split in closed form, then sums
power and transistors over that geometry plus the logic network.

Geometry explains why two arrays of equal capacity differ. What it does not
explain---the dependent logic after the read---can dominate outright:
\S\ref{sec:finding1} shows the policy with the smallest, fastest
critical-path array is nonetheless the slowest we measure, and
\S\ref{sec:latency} traces where every policy's latency goes. Two arrays with the same bit count are not the same hardware. Access
pattern fixes geometry, and geometry sets power---but the largest array is
often not the timed one, and what a policy computes \emph{after} its read can
cost more than the read itself. None of this appears in a kilobyte count.

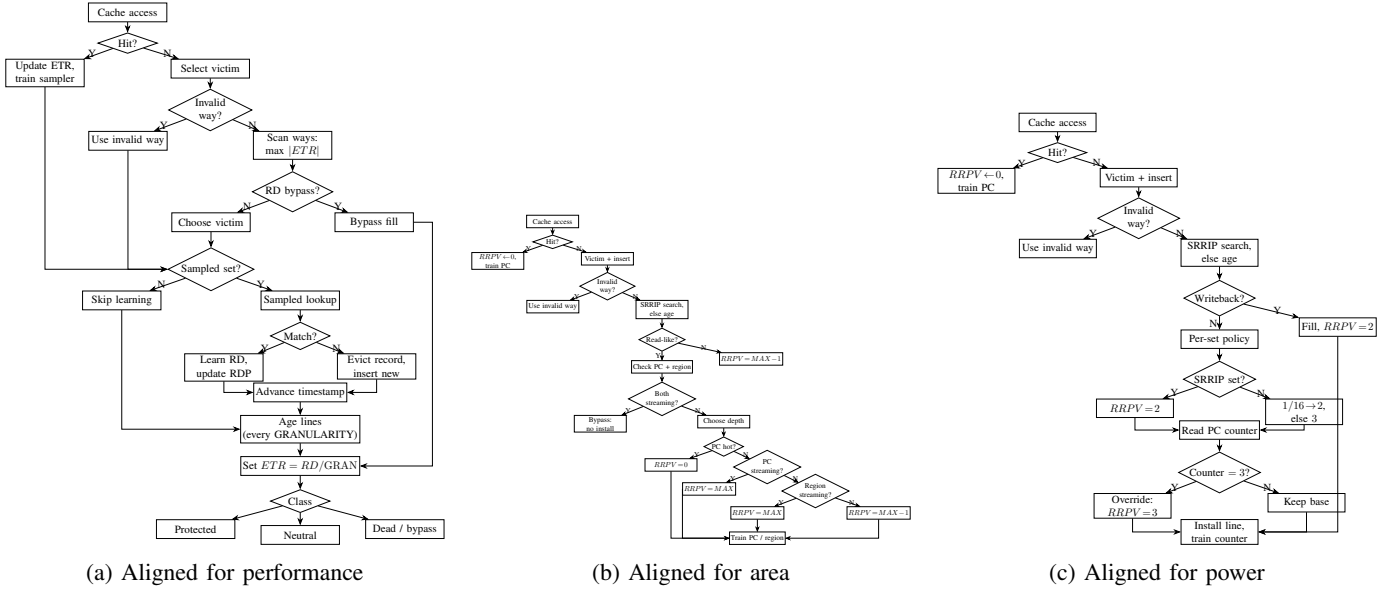
\begin{figure*}[t]
\centering

\begin{minipage}[b]{0.32\textwidth}
\centering
\resizebox{\linewidth}{!}{%
\begin{tikzpicture}[node distance=0.30cm and 0.55cm]
\path (-3cm,0) -- (3cm,0);          %

\node[cfblock] (access) {Cache access};
\node[cfdec, below=0.34cm of access] (hit) {Hit?};
\node[cfblock, below left=0.34cm and 0.9cm of hit]  (hitupd) {Update ETR,\\ train sampler};
\node[cfblock, below right=0.34cm and 0.9cm of hit] (miss)   {Select victim};
\node[cfdec, below=0.34cm of miss] (inv) {Invalid\\ way?};
\node[cfblock, below left=0.34cm and 0.7cm of inv]  (useinv) {Use invalid way};
\node[cfblock, below right=0.34cm and 0.7cm of inv] (scan)   {Scan ways:\\ max $|ETR|$};
\node[cfdec, below=0.38cm of scan] (byp) {RD bypass?};
\node[cfblock, below left=0.34cm and 0.7cm of byp]  (victim) {Choose victim};
\node[cfblock, below right=0.34cm and 0.7cm of byp] (bfill)  {Bypass fill};
\node[cfdec, below=0.50cm of victim] (samp) {Sampled set?};
\node[cfblock, below left=0.34cm and 0.9cm of samp]  (noskip) {Skip learning};
\node[cfblock, below right=0.34cm and 0.9cm of samp] (look)   {Sampled lookup};
\node[cfdec, below=0.38cm of look] (match) {Match?};
\node[cfblock, below left=0.34cm and 0.7cm of match]  (learn) {Learn RD,\\ update RDP};
\node[cfblock, below right=0.34cm and 0.7cm of match] (evict) {Evict record,\\ insert new};
\node[cfblock, below=1.05cm of match] (ts)   {Advance timestamp};
\node[cfblock, below=0.42cm of ts]    (age)  {Age lines\\ (every GRANULARITY)};
\node[cfblock, below=0.42cm of age]   (setetr) {Set $ETR = RD/\mathrm{GRAN}$};
\node[cfdec,  below=0.44cm of setetr] (cls)  {Class};
\node[cfblock, below left=0.40cm and 1.6cm of cls] (reuse) {Protected};
\node[cfblock, below=0.40cm of cls]                (neut)  {Neutral};
\node[cfblock, below right=0.40cm and 1.6cm of cls](dead)  {Dead / bypass};

\draw[cfline] (access) -- (hit);
\draw[cfline] (hit) -- node[cflab,left]  {Y} (hitupd);
\draw[cfline] (hit) -- node[cflab,right] {N} (miss);
\draw[cfline] (miss) -- (inv);
\draw[cfline] (inv) -- node[cflab,left]  {Y} (useinv);
\draw[cfline] (inv) -- node[cflab,right] {N} (scan);
\draw[cfline] (scan) -- (byp);
\draw[cfline] (byp) -- node[cflab,left]  {N} (victim);
\draw[cfline] (byp) -- node[cflab,right] {Y} (bfill);
\draw[cfline] (hitupd.south) |- (samp.west);
\draw[cfline] (useinv.south) |- (samp.west);
\draw[cfline] (victim) -- (samp);
\draw[cfline] (samp) -- node[cflab,left]  {N} (noskip);
\draw[cfline] (samp) -- node[cflab,right] {Y} (look);
\draw[cfline] (look) -- (match);
\draw[cfline] (match) -- node[cflab,left]  {Y} (learn);
\draw[cfline] (match) -- node[cflab,right] {N} (evict);
\draw[cfline] (learn.south) |- (ts.west);
\draw[cfline] (evict.south) |- (ts.east);
\draw[cfline] (noskip.south) |- (age.west);
\draw[cfline] (bfill.east) -- ++(0.6,0) |- (setetr.east);
\draw[cfline] (ts) -- (age);
\draw[cfline] (age) -- (setetr);
\draw[cfline] (setetr) -- (cls);
\draw[cfline] (cls) -- (reuse);
\draw[cfline] (cls) -- (neut);
\draw[cfline] (cls) -- (dead);
\end{tikzpicture}}
\\[0.35em]\small (a) Aligned for performance
\end{minipage}
\hfill
\begin{minipage}[b]{0.32\textwidth}
\centering
\resizebox{\linewidth}{!}{%
\begin{tikzpicture}[node distance=0.30cm and 0.55cm]
\path (-3cm,0) -- (3cm,0);

\node[cfblock] (access) {Cache access};
\node[cfdec, below=0.34cm of access] (hit) {Hit?};
\node[cfblock, below left=0.34cm and 0.9cm of hit]  (hitp) {$RRPV\!\leftarrow\!0$,\\ train PC};
\node[cfblock, below right=0.34cm and 0.9cm of hit] (miss) {Victim + insert};
\node[cfdec, below=0.34cm of miss] (inv) {Invalid\\ way?};
\node[cfblock, below left=0.34cm and 0.7cm of inv]  (useinv) {Use invalid way};
\node[cfblock, below right=0.34cm and 0.7cm of inv] (scan)   {SRRIP search,\\ else age};
\node[cfdec, below=0.44cm of scan] (rl) {Read-like?};
\node[cfblock, below right=0.34cm and 2.2cm of rl] (nob) {$RRPV\!=\!MAX\!-\!1$};
\node[cfblock, below=0.44cm of rl] (clsf) {Check PC + region};
\node[cfdec, below=0.42cm of clsf] (byp) {Both\\ streaming?};
\node[cfblock, below left=0.38cm and 0.9cm of byp]  (dobyp) {Bypass:\\ no install};
\node[cfblock, below right=0.38cm and 0.9cm of byp] (depth) {Choose depth};
\node[cfdec, below=0.42cm of depth] (pch) {PC hot?};
\node[cfblock, below left=0.34cm and 0.8cm of pch]  (mru) {$RRPV\!=\!0$};
\node[cfdec, below right=0.34cm and 0.8cm of pch]   (pcs) {PC\\ streaming?};
\node[cfblock, below left=0.34cm and 0.7cm of pcs]  (cold) {$RRPV\!=\!MAX$};
\node[cfdec, below right=0.34cm and 0.7cm of pcs]   (rgs)  {Region\\ streaming?};
\node[cfblock, below left=0.34cm and 0.7cm of rgs]  (cold2){$RRPV\!=\!MAX$};
\node[cfblock, below right=0.34cm and 0.7cm of rgs] (med)  {$RRPV\!=\!MAX\!-\!1$};
\node[cfblock, below=0.60cm of cold2] (train) {Train PC / region};

\draw[cfline] (access) -- (hit);
\draw[cfline] (hit) -- node[cflab,left]  {Y} (hitp);
\draw[cfline] (hit) -- node[cflab,right] {N} (miss);
\draw[cfline] (miss) -- (inv);
\draw[cfline] (inv) -- node[cflab,left]  {Y} (useinv);
\draw[cfline] (inv) -- node[cflab,right] {N} (scan);
\draw[cfline] (scan) -- (rl);
\draw[cfline] (rl) -- node[cflab,right] {N} (nob);
\draw[cfline] (rl) -- node[cflab,left]  {Y} (clsf);
\draw[cfline] (clsf) -- (byp);
\draw[cfline] (byp) -- node[cflab,left]  {Y} (dobyp);
\draw[cfline] (byp) -- node[cflab,right] {N} (depth);
\draw[cfline] (depth) -- (pch);
\draw[cfline] (pch) -- node[cflab,left]  {Y} (mru);
\draw[cfline] (pch) -- node[cflab,right] {N} (pcs);
\draw[cfline] (pcs) -- node[cflab,left]  {Y} (cold);
\draw[cfline] (pcs) -- node[cflab,right] {N} (rgs);
\draw[cfline] (rgs) -- node[cflab,left]  {Y} (cold2);
\draw[cfline] (rgs) -- node[cflab,right] {N} (med);
\draw[cfline] (mru.south)  |- (train.west);
\draw[cfline] (cold.west)  |- (train.west);
\draw[cfline] (cold2) -- (train);
\draw[cfline] (med.south)  |- (train.east);
\end{tikzpicture}}
\\[0.35em]\small (b) Aligned for area
\end{minipage}
\hfill
\begin{minipage}[b]{0.32\textwidth}
\centering
\resizebox{\linewidth}{!}{%
\begin{tikzpicture}[node distance=0.30cm and 0.55cm]
\path (-3cm,0) -- (3cm,0);

\node[cfblock] (access) {Cache access};
\node[cfdec, below=0.34cm of access] (hit) {Hit?};
\node[cfblock, below left=0.34cm and 0.9cm of hit]  (hitp) {$RRPV\!\leftarrow\!0$,\\ train PC};
\node[cfblock, below right=0.34cm and 0.9cm of hit] (miss) {Victim + insert};
\node[cfdec, below=0.34cm of miss] (inv) {Invalid\\ way?};
\node[cfblock, below left=0.34cm and 0.7cm of inv]  (useinv) {Use invalid way};
\node[cfblock, below right=0.34cm and 0.7cm of inv] (scan)   {SRRIP search,\\ else age};
\node[cfdec, below=0.44cm of scan] (wb) {Writeback?};
\node[cfblock, below right=0.34cm and 2.0cm of wb] (wbins) {Fill, $RRPV\!=\!2$};
\node[cfblock, below=0.44cm of wb] (base) {Per-set policy};
\node[cfdec, below=0.40cm of base] (mode) {SRRIP set?};
\node[cfblock, below left=0.34cm and 0.9cm of mode]  (srr) {$RRPV\!=\!2$};
\node[cfblock, below right=0.34cm and 0.9cm of mode] (brr) {$1/16\!\rightarrow\!2$,\\ else $3$};
\node[cfblock, below=0.75cm of mode] (pcread) {Read PC counter};
\node[cfdec, below=0.40cm of pcread] (pcs) {Counter $=3$?};
\node[cfblock, below left=0.34cm and 0.9cm of pcs]  (ovr)  {Override:\\ $RRPV\!=\!3$};
\node[cfblock, below right=0.34cm and 0.9cm of pcs] (keep) {Keep base};
\node[cfblock, below=0.85cm of pcs] (inst) {Install line,\\ train counter};

\draw[cfline] (access) -- (hit);
\draw[cfline] (hit) -- node[cflab,left]  {Y} (hitp);
\draw[cfline] (hit) -- node[cflab,right] {N} (miss);
\draw[cfline] (miss) -- (inv);
\draw[cfline] (inv) -- node[cflab,left]  {Y} (useinv);
\draw[cfline] (inv) -- node[cflab,right] {N} (scan);
\draw[cfline] (scan) -- (wb);
\draw[cfline] (wb) -- node[cflab,right] {Y} (wbins);
\draw[cfline] (wb) -- node[cflab,left]  {N} (base);
\draw[cfline] (base) -- (mode);
\draw[cfline] (mode) -- node[cflab,left]  {Y} (srr);
\draw[cfline] (mode) -- node[cflab,right] {N} (brr);
\draw[cfline] (srr.south) |- (pcread.west);
\draw[cfline] (brr.south) |- (pcread.east);
\draw[cfline] (pcread) -- (pcs);
\draw[cfline] (pcs) -- node[cflab,left]  {Y} (ovr);
\draw[cfline] (pcs) -- node[cflab,right] {N} (keep);
\draw[cfline] (ovr.south)   |- (inst.west);
\draw[cfline] (keep.south)  |- (inst.east);
\draw[cfline] (wbins.south) |- (inst.east);
\end{tikzpicture}}
\\[0.35em]\small (c) Aligned for power
\end{minipage}

\caption{\textbf{Control-flow structure of three aligned policies under different hardware constraints.}}
\label{fig:control_flow_comparison}
\end{figure*}

\medskip\subsection{Closed-loop search with an exact verifier}
\label{sec:formal}

A pattern has crystallized in the LLM-agent literature: take a pretrained
model, wrap it in a closed loop with an exact verifier, keep some memory of
past attempts, and iterate. Voyager~\cite{wang2023voyager},
Reflexion~\cite{shinn2023reflexion}, FunSearch~\cite{romera2024funsearch},
Eureka~\cite{ma2024eureka}, and AlphaEvolve~\cite{novikov2025alphaevolve} all
instantiate it, and by 2026 it has reached architecture.
CacheMind~\cite{cachemind} augments \champsim{} with retrieval-augmented
generation, answering natural-language questions about cache behavior from
trace artifacts---but it reports what a policy \emph{does}, not what it
costs, and a human drives every query. ArchAgent~\cite{archagent} and Agentic
Architect~\cite{agenticarchitect} go further and close the loop itself:
LLM-driven evolution of replacement policies (and, in the latter, prefetchers
and branch predictors) inside a cycle-accurate simulator, beating published
baselines on IPC.

Every agent above
scores a candidate with one cheap, exact number---a unit test, a game state,
IPC from a simulator run. A replacement policy's true objective is four
numbers, and three of them have never had a model to come from. Their search
is therefore free to buy speedup with per-line state, logic depth, and
power it cannot see. \sys{} enhances prior work with six interventions (\S\ref{sec:overview}) that address this gap.

In addition, large language models tend to increase the size and
structural complexity of the generated code by a phenomenon similar to
``spaghetti code,'' which prior works did not address systematically. This effect is visible in the control-flow diagrams of representative generated policies in Figure~\ref{fig:control_flow_comparison}. This growth directly impacts
the \emph{feasibility} of the policies they produce, because more
pipeline stages, more conditional logic, and more per-line state all
push against timing, area, and verification budgets. The \sys{} design
addresses this challenge by enforcing latency constraints inside the
discovery loop: victim-path latency and other stage-level budgets are
measured and fed back into the system so that overly complex,
LLM-generated policies are rejected or distilled into simpler,
mechanism-level rules. In effect, latency serves both as a proxy for
microarchitectural complexity and as a guardrail that prevents
unbounded growth of LLM-generated replacement logic.

Latency of the LLC victim path is not a first-order determinant of
end-to-end miss latency in a \emph{DRAM-dominated} regime, but it
remains important as a proxy for \textit{implementation/verification feasibility} and \emph{structural complexity}. Even when tens of picoseconds of additional logic
on the victim path are completely hidden inside a 200\,+ cycle
off-chip miss penalty, the mechanisms that create those picoseconds
(e.g., multi-stage aging chains, wide comparators, multi-level priority
encoders) still drive \emph{area}, \emph{static power}, and
\emph{timing-closure effort}. From a computer architect's perspective,
replacement policy victim-selection-path latency in the LLC is therefore a complexity signal rather
than a performance signal, and it can be used to keep the
implementation within a reasonable range of implementation cost while
the system focuses on more impactful objectives.

\section{\textsc{CacheFit}}
\subsection{System overview \& motivation}
\label{sec:sysoverview}

\sys{} is motivated by two questions. How much of a policy's hardware cost can be
removed before IPC suffers? And can that be found by hand, with the simulators
the field already has? The answer to the second is no, because the naive tuning of policies fails to arrive at the most efficient and high performance policies.

A replacement policy pays for three separate hardware quantities. State volume
sets how many bits must be built. Array organization sets how those bits are
decoded, driven, and leaked. The dependency graph sets how much work must
finish after the arrays are read. Metadata kilobytes measure only the first.
Performance depends on none of them---it depends on whether the stored state
carries useful information, not on how much state exists. The field measures the first and treats it as a proxy for all four. Table~\ref{tab:hw} shows what that substitution costs.

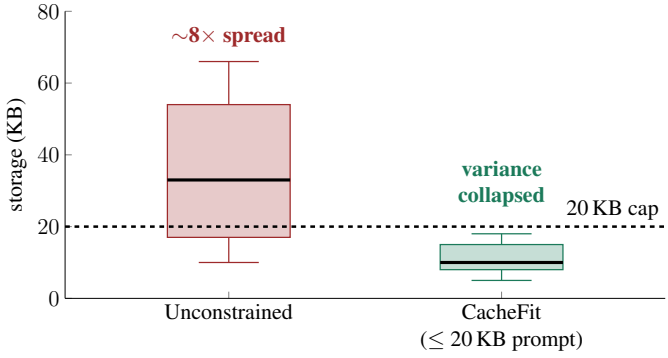
\begin{figure}[t]
\centering
\definecolor{uncred}{HTML}{9E2A2B}
\definecolor{cfgreen}{HTML}{1B7A5A}
\newcommand{\boxwhisker}[7]{%
    \draw[#7, thick] (axis cs:#1,#2) -- (axis cs:#1,#3);
    \draw[#7, thick] (axis cs:#1,#5) -- (axis cs:#1,#6);
    \draw[#7, thick] (axis cs:{#1-0.11},#2) -- (axis cs:{#1+0.11},#2);
    \draw[#7, thick] (axis cs:{#1-0.11},#6) -- (axis cs:{#1+0.11},#6);
    \draw[#7, thick, fill=#7!25]
        (axis cs:{#1-0.225},#3) -- (axis cs:{#1+0.225},#3)
        -- (axis cs:{#1+0.225},#5) -- (axis cs:{#1-0.225},#5) -- cycle;
    \draw[black, line width=2pt]
        (axis cs:{#1-0.225},#4) -- (axis cs:{#1+0.225},#4);
}
\resizebox{\columnwidth}{!}{%
\begin{tikzpicture}
\begin{axis}[
    width=15cm, height=8cm,
    ymin=0, ymax=80,
    xmin=0.4, xmax=2.6,
    ylabel={storage (KB)},
    ylabel style={font=\Large},
    ytick={0,20,40,60,80},
    yticklabel style={font=\Large},
    xtick={1,2},
    xticklabels={Unconstrained, {CacheFit\\($\leq$ 20\,KB prompt)}},
    xticklabel style={font=\Large, align=center},
    axis x line*=bottom, axis y line*=left,
    set layers,
]
\addplot[draw=none] coordinates {(1,0) (2,0)};
\pgfplotsextra{%
\begin{pgfonlayer}{main}
\boxwhisker{1}{10}{17}{33}{54}{66}{uncred}
\boxwhisker{2}{5}{8}{10}{15}{18}{cfgreen}
\end{pgfonlayer}
\begin{pgfonlayer}{axis foreground}
\draw[black, dashed, line width=1.6pt]
    (axis cs:0.4,20) -- (axis cs:2.6,20)
    node[pos=1, anchor=south east, font=\Large] {20\,KB cap};
\node[uncred,  font=\Large\bfseries] at (axis cs:1,73)
    {$\sim$8$\times$ spread};
\node[cfgreen, font=\Large\bfseries, align=center] at (axis cs:2,32)
    {variance\\collapsed};
\end{pgfonlayer}
}
\end{axis}
\end{tikzpicture}
}%
\caption{Naive search (FunSearch~\cite{romera2024funsearch} adopted for LLC replacement) vs. \sys{}  with 20\,KB constraint.}
\label{fig:variance}
\end{figure}

An unconstrained search could produce policies spanning 9 to 70\,KB of
metadata---roughly $8\times$---with no sign of settling. Figure~\ref{fig:variance}
shows that a 20\,KB cap collapses this spread into the requested band. The problem is variance, not bias: any single candidate may land several times above or below the population mean, so no amount of averaging recovers a design.

\begin{figure}[t]
\centering
\definecolor{pcsigred}{HTML}{A11D24}
\definecolor{rrpvorange}{HTML}{C46B15}
\definecolor{flagsgold}{HTML}{D8B14A}
\definecolor{shctgray}{HTML}{7A7A7A}
\definecolor{eafgreen}{HTML}{1F6B3A}
\definecolor{prctblue}{HTML}{2171B5}
\definecolor{rrpvlblue}{HTML}{6BAED6}
\resizebox{\columnwidth}{!}{%
\begin{tikzpicture}
\begin{axis}[
    ybar stacked,
    width=10cm, height=7.8cm,
    bar width=61pt,              %
    xmin=-0.5, xmax=1.5,
    ymin=0, ymax=80,
    title={(a) 70\,KB sprawl is per-line, not algorithmic},
    title style={font=\Large, yshift=-2pt},
    ylabel={storage (KB)},
    ylabel style={font=\LARGE},
    ytick={0,20,40,60,80},
    yticklabel style={font=\LARGE},
    xtick={0,1},
    xticklabels={{SHiP-Mini-\\Bypass},{EAF-RAP}},
    xticklabel style={font=\Large, align=center},
    tick align=outside,
    set layers,
    legend style={
        at={(0.98,0.96)}, anchor=north east,
        font=\normalsize, cells={anchor=west},
        draw=gray!60, fill=white, inner sep=2pt,
    },
    legend cell align=left,
    area legend,
]
\addplot[fill=pcsigred,   draw=black, line width=0.6pt] coordinates {(0,48)   (1,0)};
\addplot[fill=rrpvorange, draw=black, line width=0.6pt] coordinates {(0,8)    (1,0)};
\addplot[fill=flagsgold,  draw=black, line width=0.6pt] coordinates {(0,12)   (1,0)};
\addplot[fill=shctgray,   draw=black, line width=0.6pt] coordinates {(0,1.75) (1,0)};
\addplot[fill=eafgreen,   draw=black, line width=0.6pt] coordinates {(0,0) (1,4.5)};
\addplot[fill=prctblue,   draw=black, line width=0.6pt] coordinates {(0,0) (1,4.0)};
\addplot[fill=rrpvlblue,  draw=black, line width=0.6pt] coordinates {(0,0) (1,1.0)};
\legend{per-line PC sig (12b),
        per-line RRPV (2b),
        per-line flags (3b),
        global SHCT,
        Evicted-Address Filter,
        2-bit per-region PRCT,
        RRPV (shared, narrow)}
\pgfplotsextra{%
\begin{pgfonlayer}{axis foreground}
\node[font=\LARGE\bfseries, anchor=south] at (axis cs:0,70.5) {69.75\,KB};
\node[font=\LARGE\bfseries, anchor=south] at (axis cs:1,10.3) {9.50\,KB};
\end{pgfonlayer}
}
\end{axis}
\begin{axis}[
    at={(12.9cm,0)}, anchor=south west,
    width=9cm, height=7.8cm,
    bar width=55pt,              %
    xmin=-0.5, xmax=1.5,
    ymin=0.2350, ymax=0.2450,
    title={(b) 7$\times$ more storage, \emph{lower} IPC},
    title style={font=\Large, yshift=-2pt},
    ylabel={cumulative IPC on omnetpp},
    ylabel style={font=\LARGE},
    ytick={0.2350,0.2375,0.2400,0.2425,0.2450},
    yticklabel style={font=\LARGE, /pgf/number format/precision=4,
                      /pgf/number format/fixed, /pgf/number format/fixed zerofill},
    xtick={0,1},
    xticklabels={{SHiP-Mini-\\Bypass\\69.75\,KB},{EAF-RAP\\9.50\,KB}},
    xticklabel style={font=\Large, align=center},
    tick align=outside,
    set layers,
]
\addplot[ybar, bar shift=0pt, fill=pcsigred, draw=black, line width=0.6pt]
    coordinates {(0,0.2397)};
\addplot[ybar, bar shift=0pt, fill=eafgreen, draw=black, line width=0.6pt]
    coordinates {(1,0.2414)};
\pgfplotsextra{%
\begin{pgfonlayer}{axis foreground}
\node[font=\LARGE\bfseries, anchor=south] at (axis cs:0,0.2398) {0.2397};
\node[font=\LARGE\bfseries, anchor=south] at (axis cs:1,0.2415) {0.2414};

\draw[-stealth, semithick]
    (axis cs:0,0.2397) -- (axis cs:1,0.2414);

\draw[-stealth, semithick, shorten <=28pt]
    (axis cs:0.28,0.2399)
    -- node[draw=black, fill=white, font=\Large\bfseries, align=center,
            anchor=south, inner sep=3pt, pos=0.5, yshift=7pt]
       {+0.7\%~IPC at\\14\%~of storage}
    (axis cs:0.31,0.243);

\end{pgfonlayer}
}
\end{axis}
\end{tikzpicture}
}%
\caption{Storage breakdown and IPC of two policies from the same unguided search.}
\label{fig:teardown}
\end{figure}
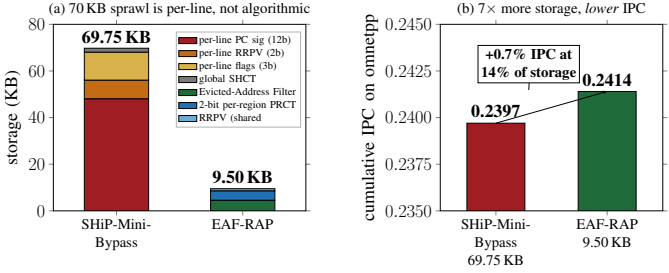

Worse, the search cannot tell when state is wasted. Two policies from the same unguided run illustrate this (Fig.~\ref{fig:teardown}). SHiP-Mini-Bypass carries
69.75\,KB and reaches IPC 0.2397; EAF-RAP carries 9.50\,KB and reaches 0.2414.
The larger policy spends an extra 17 bits on every one of 32{,}768 lines and
performs no better for it. To a generator with no cost model, per-line state
looks nearly free---but in an LLC it is the most expensive state there is,
replicated once per line. The cost constraint has to participate in the search,
not audit it afterward.

\subsection{Root Cause: Why naive Search for efficient redesign fails}
\label{sec:rootcause}

Three findings explain why a search that sees only kilobytes cannot find the
budget:

\medskip\subsubsection*{Root Cause 1: metadata is an area proxy, not a timing model}
\label{sec:finding1}
Ranked by metadata, the five directly comparable published policies run
LRU~($16.0$)~$<$~\sip{}~($20.0$)~$<$~Mockingjay~($31.9$)~$<$~MPPPB~($33.2$)~$<$~Hawkeye~($36.0$)\,KB.
Ranked by logic depth---everything that must finish after the first metadata
read---they run Hawkeye~($16^\dagger$)~$<$~LRU~($47$)~$<$~MPPPB~($197^\dagger$)~$<$~\sip{}~($286$)\\~$<$~Mockingjay~($357$)\,ps,
where $^\dagger$ marks a lower bound from a partially elaborated path. The two
orderings are close to reversed at both ends: the most expensive policy by
metadata is the cheapest in hardware, and the second-cheapest is second-most
expensive.

Static power departs for a different reason. Across the five policies it
runs from $0.0048$ to $0.0081$\,mW per kilobyte, a $1.7\times$ spread.
Fit-MPPPB makes the effect concrete against LRU: $18\%$ fewer bits, $13.2$
against $16.0$\,KB, yet $1.4\times$ the power---its state is divided among
several small arrays whose count and aspect ratios carry periphery costs no
bit total records. Transistor count supplies the control case: transistors
per kilobyte vary by only $1.04\times$, and the transistor ordering matches
the metadata ordering exactly. A one-number cost claim is defensible for
area-like cost; timing needs the dependency graph, and power needs the
array geometry.

\medskip\subsubsection*{Root Cause 2: rich prediction is cheap when it is precomputed}
\label{sec:finding2}
Hawkeye does far more algorithmic work than LRU---it reconstructs Belady's
decisions on sampled accesses and trains a reuse predictor---and almost none
of it happens when the cache must name a victim. OPTgen and training run on
sampled sets; the predictor shapes insertion after a fill; victim selection
consumes the resulting RRPV state through a single first-maximum search. LRU
trains nothing, yet every victim decision compares sixteen 4-bit values and
encodes the winner. Hawkeye's $16$\,ps is a lower bound (its full scan is not
yet elaborated), so we claim no final rank from it; the placement result
stands regardless. The distinction that matters is not simple versus
sophisticated but precomputed versus re-executed: learn richly off the demand
path, and leave one short reduction in the cycle that needs a victim.

\medskip\subsubsection*{Root Cause 3: in MPPPB, predictive value saturates before state does}
\label{sec:finding4}
Across Table~\ref{tab:hw}, aggregate IPC varies by under $1.5\%$ while
hardware cost varies by multiples. \sip{} at $20$\,KB outperforms Hawkeye at
$36.0$\,KB; MPPPB at $33.2$\,KB delivers $0.76\%$ \emph{less} IPC than LRU.
Two saturation regimes make this unsurprising at a $2$\,MB LLC: when the
working set fits, replacement has few mistakes left to prevent; when accesses
stream, no predictor state manufactures reuse. Fit-MPPPB turns the
observation into a within-policy experiment---storage $33.2 \to 13.2$\,KB,
power $0.270 \to 0.118$\,mW, transistors $1778 \to 730$\,K---and IPC
\emph{rises}, $0.6107 \to 0.6174$. Most of the removed state had no marginal
predictive value, and the ensemble structure explains why the reduction can
be gradual: MPPPB sums independently indexed votes, so table count, counter
width, and sampler shape turn continuously, where shrinking a single-table
predictor aliases histories and one collision can erase the only observation
a decision had. We establish this on the policies tested, not as a law; the
testable hypothesis is that independently indexed ensembles expose smoother
cost controls than single-table predictors.

Together the findings show the four quantities of \S\ref{sec:sysoverview} move
independently: a larger policy can have a shorter online path, equal bit
counts can leak differently, and removing predictor state can raise IPC. A
search that optimizes kilobytes optimizes one of the four. A circuit-aware
search can resize state, reshape arrays, and move work off the victim path,
measuring IPC after every change---the problem \sys{} solves.

\subsection{AI-assisted design exploration \& novel features compared to naive LLM loop}
\label{sec:overview}

\sys{} is a training-free, closed-loop framework that discovers cache
replacement policies as executable code. Three components: an LLM generator
that writes complete \champsim{}-compatible policies under the cache
interface, hardware budgets, and retrieved design lessons; an evaluation
backend that compiles each candidate, measures IPC in \champsim{}, and costs
it in \harcom{} (latency, storage, area, power, transistors, with per-stage
attribution); and a controller that gates on feasibility and distills measured
outcomes into persistent, editable memory. Each iteration runs \emph{propose
$\rightarrow$ compile $\rightarrow$ simulate $\rightarrow$ cost $\rightarrow$
gate $\rightarrow$ distill $\rightarrow$ retrieve}. Compile errors trigger
bounded repair; any candidate violating a hardware constraint is rejected even
when it improves IPC. The distilled memory follows CacheMind~\cite{cachemind}
in grounding natural-language claims about cache behavior in measured
artifacts rather than model recall---here the artifact is a \harcom{} report
and a \champsim{} run rather than a trace slice. Because that memory lives
outside the model, an architect can inspect, edit, or delete any rule without
retraining.

\medskip\subsubsection*{Feature 1: memory integration}
\label{sec:method}
\label{sec:memvars}
There are three memory configurations, differing only in $\pi$ and $U$.
\textit{No memory}: each prompt carries the system contract and the
best-so-far parent, nothing else. \textit{Replay memory}: an archive
stores every simulated policy; iteration $t$ pastes the top-$n$ by IPC
into the prompt as examples---comparable to experience replay in
RL~\cite{lin1992replay} and Reflexion-style trajectory
buffers~\cite{shinn2023reflexion}; richer
schemes~\cite{wang2023voyager,romera2024funsearch,ma2024eureka} all
contain this top-$n$-by-score component. \textit{Summary memory}: four
typed fields (\texttt{PRINCIPLES}, \texttt{DONTS}, \texttt{KNOBS},
\texttt{OPEN-Q}), $32$ rows maximum, written by an LLM distillation step
after each simulation and evicted by recency-weighted relevance; every
entry carries provenance to its experiment, and no entry is itself a
candidate design.

\noindent{\textit{From Raw History to Summary Memory:}}\label{sec:int1}
Raw history re-anchors generation on the best archetypes already
found---the model sees winning artifacts, code and score, and imitates
them. Summary memory persists mechanism-level lessons instead; capping
it at $32$ rows constrains the channel between past samples and the next
prompt in the information-bottleneck sense~\cite{tishby1999bottleneck}:
the loop forgets the archetype and keeps the lesson. Distillation alone
is not exploration---the remaining features restore the diversity it
cannot.

\medskip\subsubsection*{Feature 2: stagnation-triggered crossover}
\label{sec:int3}

Single-parent refinement is local---the proposal distribution has support near
the current best. On stagnation, the controller instead draws two parents from
the top-$m$ archive ($m=8$) and prompts for a child retaining identifiable
blocks from both. The child faces the same \champsim{} pipeline as any
candidate; surviving, it becomes the new lineage root. Random second-parent
selection is too undirected, which Feature~3 fixes.

\medskip\subsubsection*{Feature 3: distance-based parent selection}
\label{sec:int4}

Quality-diversity search~\cite{mouret2015quality,cully2015robots} structures its
archive by behavioral diversity and favors distant parents. Feature~3 chooses Parent~$B$ to maximize embedding distance from Parent~$A$,
\begin{equation}
B \;=\; \arg\max_{P \in T_m \setminus \{A\}}\,\bigl(1 - \cos(e(A), e(P))\bigr),
\end{equation}
where $e(\cdot)$ is the MiniLM~\cite{minilm2020} embedding of the C++ policy
(\texttt{all-MiniLM-L6-v2}). Random selection from a score-anchored archive
returns near-duplicates of $A$ in expectation; max-distance selection forces
the model to integrate genuinely different mechanisms. Our M2 leaderboard
policy (IPC $0.41643$) is representative: produced when refinement plateaued at
iteration~$22$, it hybridizes a bypass-heavy parent with a recency- and
prefetch-aware one and wins on \texttt{lbm}, \texttt{mcf}, \texttt{namd}, and
\texttt{zeusmp}, where neither parent wins alone.

\medskip\subsubsection*{Feature 4: auditable, editable memory feature} 
\label{sec:int5}
Editing $M_t$ changes the proposal distribution, and three primitives suffice.
\emph{Principle injection} inserts a rule the loop has not derived.
\emph{Forced forgetting} deletes every row mentioning a mechanism---%
\texttt{RRPV}, say---to reopen a region the loop had closed. \emph{Lesson
rewriting} downgrades a \texttt{DONTS} row to \texttt{KNOBS} or
\texttt{OPEN-Q} when distillation has overgeneralized from one attribution.

\medskip\subsubsection*{Feature 5: foundation-model rotation and stagnation gate}
\label{sec:int6}

\sys{} also rotates a pool of foundation models $\mathcal{B} =
\{$\texttt{gpt-oss-20b}, \texttt{Qwen3-Coder}, 
\texttt{o4-mini}$\}$~\cite{gptoss2024,qwen2024},
excluding on stagnation whichever backbone produced the most canon-like recent
outputs.

One trigger drives all of this. Let $S_i$ be the composite reward at iteration
$i$ and $\Delta_i = \max(0, S_i - \max_{j<i} S_j)$. A budget $b$ starts at $5$,
gains $2$ when $\Delta_i > \epsilon$, and loses $1$ otherwise. At $b \le 0$ the
run is stagnated: the next iteration considers crossover, rotates the backbone,
and rebuilds the summary from recent traces. Without the shared gate, the
interventions fire independently and thrash.

\medskip\subsubsection*{Feature 6: cost integration}
\label{sec:harcom}
Each candidate produced by the search loop is evaluated on
two independent axes. Its \code{.cc} is compiled and run against real
SPEC2017 traces in ChampSim, giving measured IPC and hit rate. In parallel,
the same policy is translated into a \harcom{} header --- every declared
metadata array becomes a \code{ram} of explicit depth and word width, and
every inter-array operation becomes a typed \harcom{} primitive --- which is
compiled and run to obtain real storage, transistor count, area, static
power, and latency. Translation is retried against both compiler errors and
a mechanical fidelity check that flags shortcuts such as collapsing
non-uniform tables into a uniform fallback.

Total victim-selection latency $t_\text{total}$ is measured
directly: the compiled \harcom{} binary calls \code{get\_victim} once and
times the whole circuit from input to output. $t_\text{sram}$ is not measured the same way --- it comes from a
separate evaluation of \harcom{}'s closed-form SRAM model
(Eq.~\ref{eq:sram}) applied to whichever array \code{get\_victim} actually
reads, computed purely from that array's row and column counts, with no
circuit execution needed. This distinction matters because a policy can
declare several metadata structures, and only the one on
\code{get\_victim}'s own read path sets $t_\text{sram}$; a differently-shaped
array touched only during \code{update\_replacement\_state} is off the
timed path and does not belong in the comparison. Logic overhead is then the
difference, $t_\text{logic} = t_\text{total} - t_\text{sram}$: whatever
combinational work --- aging, reduction, priority encoding --- the policy
does after the read and before a victim is decided.

\medskip\section{Experimental Setup}
\label{sec:expsetup}

We evaluate every policy in \champsim{} CRC2~\cite{champsim}, single-core, 2\,MB/2048-set/16-way/64\,B LLC matching the \harcom{}
target, private non-inclusive L1/L2, with prefetching, single
64-bit FR-FCFS DRAM channel ($\approx$13.5/40.5\,ns row-hit/miss), 100\,M
warmup + 1B instructions. Full parameters are in Table~\ref{tab:simconfig-app}. 

We use 20 SPEC~CPU2017 --which we could compile their simpoint without errors.
The six published
baselines---LRU, \hawkeye{}~\cite{hawkeye}, LIME~\cite{lime}, RED~\cite{red},
\sip{}~\cite{shippp}, \mpppb{}~\cite{mpppb}---are compiled from the
\champsim{} CRC2 policy library.

\begin{table}[!htbp]
\renewcommand{\arraystretch}{1.05}
\caption{Simulation configuration.}
\label{tab:simconfig-app}
\centering
\small
\setlength{\tabcolsep}{3pt}
\begin{tabular}{@{}>{\raggedright\arraybackslash}p{1.35cm}
                  >{\raggedright\arraybackslash}p{\dimexpr\columnwidth-1.35cm-6pt\relax}@{}}
\toprule
\multicolumn{2}{@{}l}{\textit{Core and cache hierarchy}} \\
Core      & ChampSim CRC2~\cite{champsim}, \texttt{config2.a}; 6-wide OoO,
             256-entry ROB, GShare \\
L1D / L2  & 32\,KB, 8-way, 4-cyc / 256\,KB, 8-way, $+8$-cyc \\
LLC       & 2\,MB, 2048 sets $\times$ 16 ways, 64\,B, $+20$-cyc, non-inclusive \\
Prefetch  & Next-line (L1D), PC-stride (L2) \\
DRAM      & 1$\times$64-bit, FR-FCFS; row hit/miss $\approx$13.5/40.5\,ns \\
\midrule
\multicolumn{2}{@{}l}{\textit{Workloads}} \\
Suite     & SPEC CPU 2017 Speed, 20 benchmarks (12 int, 8 fp) \\
SimPoint  & One top-weight SimPoint~3.2~\cite{simpoint} per benchmark;
             BBVs via Valgrind \texttt{exp-bbv}~3.11.0 \\
Interval  & 100\,M warmup $+$ 1B instructions \\
Metrics   & IPC harmonic mean, hit rate arithmetic mean (20 workloads) \\
\midrule
\multicolumn{2}{@{}l}{\textit{\harcom{} hardware model~\cite{harcom}}} \\
Process   & 5\,nm FinFET, $V_\text{DD}=0.75$\,V; 6T SRAM cell,
             $0.02\,\mu\text{m}^2$ \\
Devices   & $I_\text{dsat}$ 60/40\,$\mu$A/fin, $I_\text{off}$ 1\,nA/17\,pA per
             fin (logic/SRAM) \\
Wires     & 0.2\,fF/$\mu$m metal cap.; DSE 6 \\
\bottomrule
\end{tabular}
\end{table}

\section{Results}
\label{sec:results}

\subsection{Per-workload results}
\label{sec:mpppb-perwl}

Fit-MPPPB matches or beats baseline MPPPB's hit rate on 19 of 20 workloads
(improving on 18) and its IPC on 16 of 20 (improving on 14). The aggregate gain is highest for \code{cactuBSSN} ($+44.6\%$ IPC and $+21.3$ hit-rate points), while each of the four regressions is below $2.3\%$. Reporting the full table
makes that concentration visible rather than allowing the harmonic mean to hide it.

\begin{figure}[!htbp]
\centering
\begin{tikzpicture}
\begin{axis}[
    width=\columnwidth, height=0.75\columnwidth,
    font=\small,
    xmin=-0.5, xmax=19.5, ymin=0, ymax=54,
    xtick={0,1,...,19},
    xticklabels={imagick\_s,exchange2\_s,x264\_s,bwaves\_s,pop2\_s,roms\_s,xz\_s,
                 fotonik3d\_s,cam4\_s,cactuBSSN\_s,perlbench\_s,nab\_s,wrf\_s,
                 deepsjeng\_s,lbm\_s,gcc\_s,leela\_s,xalancbmk\_s,mcf\_s,omnetpp\_s},
    x tick label style={rotate=45, anchor=east, font=\footnotesize},
    ylabel={IPC / mW},
    xlabel={SPEC CPU2017 workload},
    tick align=inside,
    xtick pos=bottom, ytick pos=left,
    axis line style={black, line width=0.9pt},
    every axis plot/.append style={black, line width=1pt, mark size=2.6pt},
    legend style={
        at={(0.5,1.03)}, anchor=south,
        legend columns=3,
        draw=black, font=\footnotesize,
        column sep=6pt,
        /tikz/every even column/.append style={column sep=8pt},
    },
    legend cell align=left,
]

\addplot[name path=lembas, black, line width=1.6pt, mark=*, mark options={fill=black}]
  coordinates {(0,52.18)(1,32.80)(2,30.51)(3,25.16)(4,24.93)(5,21.05)(6,20.50)
  (7,19.35)(8,17.95)(9,17.80)(10,16.66)(11,15.60)(12,13.27)(13,13.02)(14,10.04)
  (15,9.35)(16,9.31)(17,8.03)(18,6.68)(19,5.49)};
\addlegendentry{Fit-SHiP++ (13.62)}

\addplot[name path=lru, black, dashed, mark=square, mark options={solid, fill=white}]
  coordinates {(0,28.14)(1,17.68)(2,16.47)(3,13.57)(4,13.74)(5,11.35)(6,11.04)
  (7,10.14)(8,9.63)(9,12.62)(10,8.98)(11,8.41)(12,7.16)(13,7.02)(14,5.29)
  (15,5.02)(16,5.02)(17,4.34)(18,3.75)(19,2.97)};
\addlegendentry{LRU (7.42)}

\addplot[gray!20, forget plot] fill between[of=lembas and lru];

\addplot[black, densely dotted, mark=triangle, mark options={solid, fill=white}]
  coordinates {(0,15.65)(1,9.84)(2,9.17)(3,7.55)(4,7.59)(5,6.31)(6,6.18)(7,5.82)
  (8,5.42)(9,6.89)(10,5.00)(11,4.68)(12,3.98)(13,3.90)(14,2.95)(15,2.82)(16,2.79)
  (17,2.41)(18,2.10)(19,1.67)};
\addlegendentry{SHiP++ (4.15)}

\addplot[black, dashdotted, mark=diamond, mark options={solid, fill=white}]
  coordinates {(0,13.48)(1,8.47)(2,7.90)(3,6.50)(4,6.51)(5,5.44)(6,5.25)(7,4.82)
  (8,4.68)(9,6.04)(10,4.30)(11,4.03)(12,3.43)(13,3.36)(14,2.46)(15,2.42)(16,2.41)
  (17,2.08)(18,1.79)(19,1.44)};
\addlegendentry{Hawkeye (3.55)}

\addplot[black, densely dashed, mark=triangle,
         mark options={solid, rotate=180, fill=white}]
  coordinates {(0,9.01)(1,5.66)(2,5.23)(3,4.34)(4,4.31)(5,3.63)(6,3.56)(7,3.51)
  (8,3.12)(9,2.71)(10,2.87)(11,2.69)(12,2.29)(13,2.24)(14,1.74)(15,1.62)(16,1.60)
  (17,1.39)(18,1.20)(19,0.96)};
\addlegendentry{MPPPB (2.36)}

\end{axis}
\end{tikzpicture}
\caption{Sorted per-workload IPC/mW for Fit-SHiP++ and baselines across the
20 SPEC~CPU2017 workloads. Legend values are aggregate IPC/mW.}
\label{fig:ipc-per-mw-workload}
\end{figure}
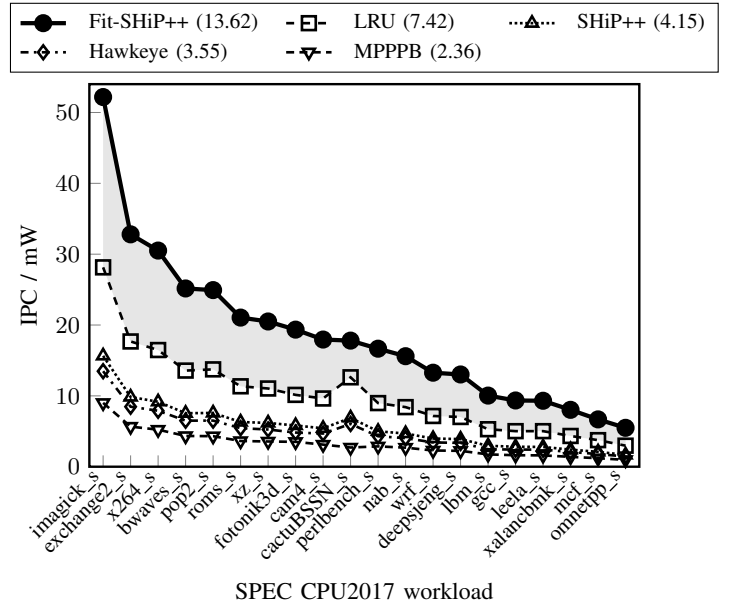

Consistent with published SPEC CPU2017 characterizations~\cite{limaye2018ispass,
singh2019icpe}, CacheFit helps the memory intensive workloads with large working sets for the LLC but high reuse---\code{cactuBSSN\_s}
most of all, then \code{roms\_s}, \code{wrf\_s}, \code{leela\_s}, \code{perlbench\_s}, \code{x264\_s}, \code{pop2\_s} and \code{cam4\_s}. It leaves the rest unchanged: workloads that already fit in cache
(\code{exchange2\_s}, \code{imagick\_s}), those that stream data without reuse
(\code{bwaves\_s}, \code{lbm\_s}), and those whose data is far too large to
retain (\code{mcf\_s}, \code{omnetpp\_s}, \code{deepsjeng\_s}).

\begin{figure}[!htbp]
\centering
\begin{tikzpicture}
\begin{axis}[
    width=\columnwidth, height=0.75\columnwidth,
    ymin=0, ymax=112,
    xmin=-0.6, xmax=19.4,
    ylabel={LLC miss rate (\%)},
    xlabel={SPEC CPU2017 workload},
    xtick=data,
    xticklabels={%
      exchange2\_s,bwaves\_s,gcc\_s,nab\_s,
      deepsjeng\_s,perlbench\_s,fotonik3d\_s,cam4\_s,
      x264\_s,roms\_s,lbm\_s,wrf\_s,
      omnetpp\_s,pop2\_s,leela\_s,xz\_s,
      mcf\_s,cactuBSSN\_s,xalancbmk\_s,imagick\_s},
    xticklabel style={rotate=45, anchor=north east, font=\scriptsize},
    yticklabel style={font=\scriptsize},
    label style={font=\small},
    ytick={0,20,40,60,80,100},
    minor y tick num=1,
    tick align=inside,
    tick pos=both,
    axis line style={black, line width=0.7pt},
    legend style={at={(0.03,0.03)}, anchor=south west, draw=black,
                  fill=white, font=\scriptsize, row sep=1pt},
    legend cell align=left,
]
\addplot[black, line width=0.8pt, mark=*, mark size=1.6pt,
         mark options={fill=black, draw=black}] coordinates {
    (0,100.000)  (1,99.724)  (2,91.416)  (3,90.657)
    (4,83.130)   (5,96.182)  (6,81.842)  (7,85.744)
    (8,89.518)   (9,99.586)  (10,72.054) (11,92.275)
    (12,63.255)  (13,67.581) (14,75.672) (15,47.652)
    (16,45.831)  (17,29.700) (18,6.492)  (19,1.461)
};
\addlegendentry{MPPPB, Mean $=70.99$\%}

\addplot[black, dashed, line width=0.8pt, mark=square, mark size=1.8pt,
         mark options={solid, fill=white, draw=black, line width=0.7pt}]
coordinates {
    (0,100.000)  (1,99.116)  (2,90.922)  (3,88.071)
    (4,86.131)   (5,81.509)  (6,81.096)  (7,79.188)
    (8,76.808)   (9,74.783)  (10,71.204) (11,69.288)
    (12,61.680)  (13,56.858) (14,55.610) (15,41.614)
    (16,40.878)  (17,8.436)  (18,4.385)  (19,1.460)
};
\addlegendentry{Fit-MPPPB, Mean $=63.45$\%}
\end{axis}
\end{tikzpicture}
\caption{Sorted per-workload LLC miss rate for Fit-MPPPB and baseline MPPPB across the 20 SPEC CPU2017 workloads.}
\label{fig:fit-mpppb-missrate}
\end{figure}
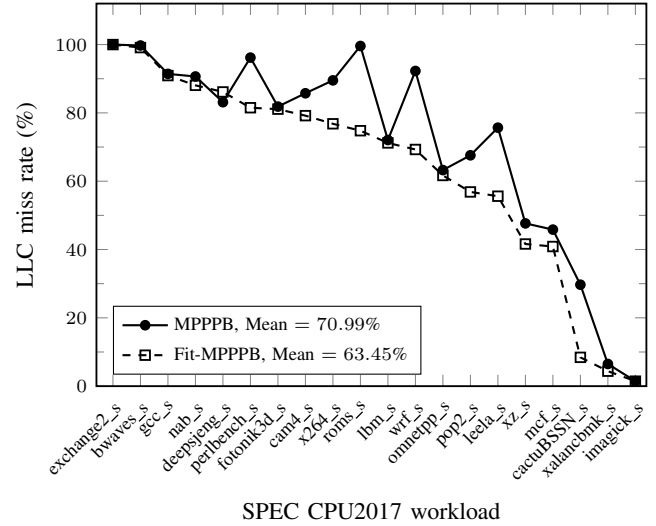

\begin{figure}[!htbp]
\centering

\begin{tikzpicture}
\begin{axis}[
    ybar stacked,
    width=\columnwidth,
    height=3.8cm,
    bar width=13pt,
    ymin=0, ymax=100,
    ytick={0,20,40,60,80,100},
    yticklabel={\pgfmathprintnumber{\tick}\,\%},
    ymajorgrids=true,
    grid style={dotted,gray!40},
    symbolic x coords={SHiP++,ReD,Hawkeye,MPPPB,Mockingjay,CacheFit},
    xtick=data,
    xticklabel style={rotate=35,anchor=east,font=\footnotesize},
    tick label style={font=\footnotesize},
    enlarge x limits=0.10,
    legend style={
        font=\footnotesize,
        at={(0.5,1.08)},
        anchor=south,
        legend columns=-1,
        draw=none
    },
]

\addplot+[
    ybar,
    draw=black,
    fill=black!20,
    pattern=north east lines
] coordinates {
    (SHiP++,22.2) (ReD,11.1) (Hawkeye,14.8)
    (MPPPB,33.3) (Mockingjay,20.0) (CacheFit,25.9)
};
\addlegendentry{WORSE\_LRU}

\addplot+[
    ybar,
    draw=black,
    fill=gray!55
] coordinates {
    (SHiP++,0.0) (ReD,14.8) (Hawkeye,14.8)
    (MPPPB,3.7) (Mockingjay,50.0) (CacheFit,0.0)
};
\addlegendentry{SAME\_LRU}

\addplot+[
    ybar,
    draw=black,
    fill=white,
    nodes near coords,
    point meta=explicit symbolic,
    every node near coord/.append style={
        font=\footnotesize,
        fill=white,
        fill opacity=0.85,
        text opacity=1,
        inner sep=1pt,
        yshift=-9pt
    }
] coordinates {
    (SHiP++,77.8)     [77.8]
    (ReD,74.1)        [74.1]
    (Hawkeye,70.4)    [70.4]
    (MPPPB,63.0)      [63]
    (Mockingjay,30.0) [30]
    (CacheFit,74.1)   [74.1]
};
\addlegendentry{BETTER\_LRU}

\end{axis}
\end{tikzpicture}

\begin{tikzpicture}
\begin{axis}[
    ybar stacked,
    width=\columnwidth,
    height=3.6cm,
    bar width=12pt,
    ymin=0, ymax=100,
    ymajorgrids=true,
    grid style={dotted,gray!40},
    ytick={0,20,40,60,80,100},
    yticklabel={\pgfmathprintnumber{\tick}\,\%},
    symbolic x coords={SHiP++,ReD,Hawkeye,MPPPB,Mockingjay,CacheFit},
    xtick=data,
    xticklabel style={rotate=45,anchor=east,font=\footnotesize},
    tick label style={font=\footnotesize},
    enlarge x limits=0.12,
    nodes near coords,
    every node near coord/.append style={
      font=\footnotesize,
      draw=none,
      fill=white,
      fill opacity=1,
      text opacity=1,
      inner sep=1pt,
    },
    nodes near coords align={vertical},
]

\addplot+[ybar, draw=black, fill=black!20, point meta=explicit,
          pattern=north east lines] coordinates {
    (SHiP++,100) [100] (ReD,100) [100] (Hawkeye,100) [100]
    (MPPPB,100) [100] (Mockingjay,100) [100] (CacheFit,0) []
};

\addplot+[ybar, draw=black, fill=gray!55] coordinates {
    (SHiP++,0) (ReD,0) (Hawkeye,0)
    (MPPPB,0) (Mockingjay,0) (CacheFit,0)
};

\addplot+[ybar, draw=black, fill=white, point meta=explicit] coordinates {
    (SHiP++,0) [] (ReD,0) [] (Hawkeye,0) []
    (MPPPB,0) [] (Mockingjay,0) [] (CacheFit,100) [100]
};

\end{axis}
\end{tikzpicture}

\begin{tikzpicture}
\begin{axis}[
    ybar stacked,
    width=\columnwidth,
    height=3.6cm,
    bar width=12pt,
    ymin=0, ymax=100,
    ymajorgrids=true,
    grid style={dotted,gray!40},
    ytick={0,20,40,60,80,100},
    yticklabel={\pgfmathprintnumber{\tick}\,\%},
    symbolic x coords={SHiP++,ReD,Hawkeye,MPPPB,Mockingjay,CacheFit},
    xtick=data,
    xticklabel style={rotate=45,anchor=east,font=\footnotesize},
    tick label style={font=\footnotesize},
    enlarge x limits=0.12,
    nodes near coords,
    every node near coord/.append style={
      font=\footnotesize,
      draw=none,
      fill=white,
      fill opacity=1,
      text opacity=1,
      inner sep=1pt,
    },
    nodes near coords align={vertical},
]

\addplot+[ybar, draw=black, fill=black!20, point meta=explicit,
          pattern=north east lines] coordinates {
    (SHiP++,100) [100] (ReD,100) [100] (Hawkeye,100) [100]
    (MPPPB,100) [100] (Mockingjay,100) [100] (CacheFit,0) []
};

\addplot+[ybar, draw=black, fill=gray!55] coordinates {
    (SHiP++,0) (ReD,0) (Hawkeye,0)
    (MPPPB,0) (Mockingjay,0) (CacheFit,0)
};

\addplot+[ybar, draw=black, fill=white, point meta=explicit] coordinates {
    (SHiP++,0) [] (ReD,0) [] (Hawkeye,0) []
    (MPPPB,0) [] (Mockingjay,0) [] (CacheFit,100) [100]
};

\end{axis}
\end{tikzpicture}

\caption{Fraction of workloads where each policy is worse than, similar to,
or better than LRU, in terms of IPC values (top), in terms of IPC/KB
(middle), and in terms of IPC/mW (bottom).}
\label{fig:lru_fraction}
\end{figure}
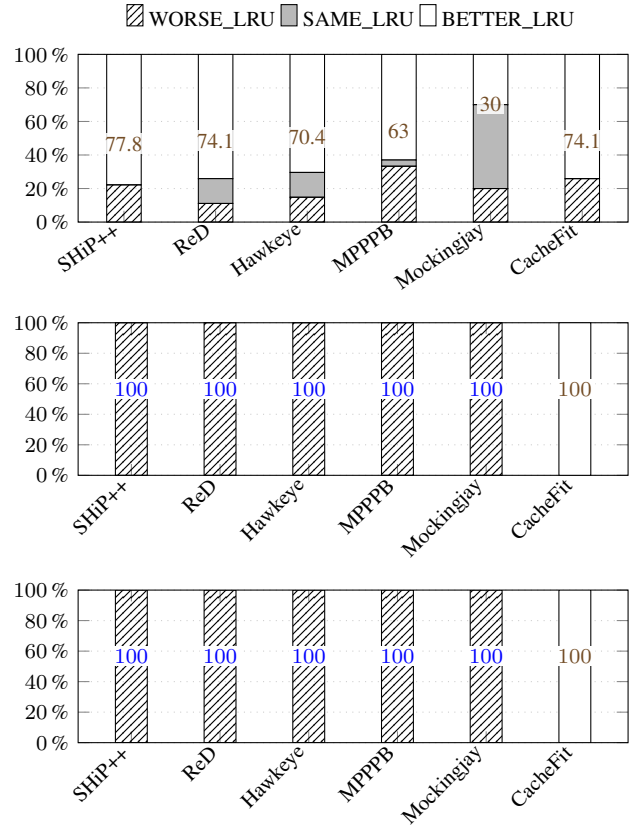

Fit-MPPPB leads baseline MPPPB on every workload and by about $2.6\times$ under
both aggregate conventions: $6.39$ versus $2.49$ for the mean of per-workload
ratios, and $4.68$ versus $1.84$ for the ratio of aggregates. We report both
because they answer different questions, although the ranking is unchanged.
LRU still has the best storage-normalized density among the published baselines.
Fit-MPPPB is different because it improves both IPC and storage relative to the
original MPPPB rather than trading one for the other. Fig.~\ref{fig:lru_fraction} shows the per-workload version of this
claim: baselines that beat LRU on raw IPC lose to it on every workload once
IPC/KB or IPC/mW is scored, and \sys{} is the only policy ahead of LRU on
all three axes.

Figure~\ref{fig:fit-mpppb-missrate} reports per-workload LLC miss rate for Fit-MPPPB and the baseline MPPPB, with workloads ordered by descending miss rate under Fit-MPPPB. Fit-MPPPB lowers the aggregate miss rate from
70.99\% to 63.45\%, a reduction of 7.54 percentage points, and is lower on 18
of the 20 workloads. The largest reductions occur on \code{roms\_s} (24.8~pp),
\code{wrf\_s} (23.0~pp), \code{cactuBSSN\_s} (21.3~pp), and \code{leela\_s}
(20.1~pp), while \code{deepsjeng\_s} is the sole regression at $3.0$
percentage points.

Figure~\ref{fig:ipc-per-mw-workload} reports per-workload power efficiency (IPC/mW) for Fit-SHiP++ against the four baselines across the SPEC~CPU2017 workloads, sorted by Fit-SHiP++'s efficiency. Fit-SHiP++ attains the highest IPC/mW on every workload, and the policy ordering (Fit-SHiP++ $>$ LRU $>$ SHiP++ $>$ Hawkeye $>$ MPPPB) never inverts, yielding an aggregate of $13.62$ IPC/mW against $7.42$ for the strongest baseline, LRU--a $1.8\times$
improvement. The margin is widest on cache-resident workloads such as \texttt{imagick\_s} ($52.2$ vs.\ $28.1$) and narrows on memory-bound ones like \texttt{omnetpp\_s} ($5.5$ vs.\ $3.0$), but Fit-SHiP++ leads throughout. 

\medskip\subsection{Generalization beyond the trained workloads}

In a separate experiment we test whether a policy tuned on a subset of workloads holds its advantage across a full benchmark suite.
Perf-SHiP++ (policy~008) was trained on nine memory-intensive SPEC~CPU2006 workloads and evaluated on the twenty-seven-workload suite. 

Perf-SHiP++ leads all six baseline policies on both sets
(Table~\ref{tab:perfshippp-generalization}). On the trained workloads it reaches a harmonic-mean IPC of $0.372$, $+5.86\%$ over LRU and $+1.14\%$ over SHiP++,
the strongest baseline in our suite. Across the full suite it reaches $0.496$, $+3.40\%$ over LRU and $+0.39\%$ over SHiP++, and stays ahead of Hawkeye,
Multiperspective, and the reordering-based policy by $1.35\%$, $0.88\%$, and $1.93\%$ respectively. 
Training on a memory-intensive slice therefore does not cost performance elsewhere.

\begin{table}[t]
\centering
\small
\begin{tabular}{lcccc}
\toprule
& \multicolumn{2}{c}{Trained (9)} & \multicolumn{2}{c}{Full suite (27)} \\
\cmidrule(lr){2-3}\cmidrule(lr){4-5}
Policy & IPC & $\Delta$\% & IPC & $\Delta$\% \\
\midrule
LRU                  & 0.351 & --    & 0.480 & --    \\
SHiP++               & 0.368 & +4.67 & 0.494 & +3.00 \\
Hawkeye              & 0.361 & +2.76 & 0.489 & +2.03 \\
Multiperspective     & 0.365 & +3.74 & 0.492 & +2.50 \\
Less is More         & 0.360 & +2.40 & 0.487 & +1.61 \\
Reordering-based     & 0.357 & +1.45 & 0.486 & +1.44 \\
\midrule
\textbf{Perf-SHiP++} & \textbf{0.372} & \textbf{+5.86} & \textbf{0.496} & \textbf{+3.40} \\
\bottomrule
\end{tabular}
\caption{Harmonic-mean IPC on SPEC~CPU2006 for \textbf{Perf-SHiP++} and six
baselines, on the nine workloads used for tuning and on the full
twenty-seven-workload suite. $\Delta\%$ is relative to LRU.}
\label{tab:perfshippp-generalization}
\end{table}

\medskip\subsection{Where the Latency Goes}
\label{sec:latency}

Table~\ref{tab:hw} separates metadata-read delay from the work that follows
it. The ordering is not a publication-year trend; it is a map of what each
translated hot path touches. The cases below show three distinct limits: wide
or tall arrays, dependent post-read logic, and useful work placed after the
victim is already known.

LRU is array-bound. A 4-bit recency counter per way per set is a single
$\langle2048,64\rangle$ RAM. \code{GetVictimInSet} reads the row ($269$\,ps),
builds a one-hot mask of maximum-recency ways, and priority-encodes it
($47$\,ps). LRU is $85\%$ array-bound, and the $64$-bit word is why. RRIP's
contribution~\cite{rrip} is precisely the observation that $4$ bits of recency
precision is far more than a recency \emph{order} needs.

Hawkeye is tall and shallow. Victim selection reads a flat
$\langle32768,3\rangle$ RRPV array. The array is sixteen times taller than
LRU's, so its row-decoder term ($30\%$ of the $268$\,ps read) nearly cancels the
benefit of the narrower word---an instance of Eq.~\ref{eq:sram} trading
$t_\text{WLSEL}$ against $t_\text{WLRC}$. Logic overhead is $16$\,ps, a lower
bound. %

\medskip\subsection{Fit-MPPPB: The Search Applied to the Most Expensive Baseline}
\label{sec:mpppb-xs8}

We explain one policy from our search pointed at MPPPB: the most expensive policy we
measure on power and, at $33.2$\,KB, on single-core storage. The objective
caps victim-path latency at MPPPB's own $284$\,ps and minimizes storage and
static power jointly, with IPC maximized subject to the cap.

The search converges on three moves: halve the predictor-table count
($16\to8$), narrow the weight counters ($6$-bit $\to$ $5$-bit), and shrink the
sampler ($80\!\times\!18 \to 64\!\times\!16$). All three were confirmed in
the executed configuration branch rather than inferred from the candidate's
description (\S\ref{sec:method}). The name records
the two visible ones: Fit-MPPPB is an \emph{extra-small} MPPPB with
\emph{eight} tables.

\noindent\harcom{} on the resulting header (Table~\ref{tab:hw}) reports $107{,}881$ bits
($13.2$\,KB, $-60.3\%$), $0.118$\,mW static power ($-56.3\%$), and $-58.9\%$
transistors. On the full SPEC2017 harness IPC \emph{improves}
($0.6174$ against $0.6107$, $+1.1\%$ aggregate), LLC hit rate rises from
$0.290$ to $0.365$, and misses per thousand instructions fall $15\%$. Cutting
storage and power by more than half cost nothing on either performance axis we
measure.

\ins{}{Halving the predictor did not halve its decision quality: Fit-MPPPB matches
or beats baseline MPPPB's hit rate on 19 of 20 workloads with $40\%$ of the
storage.}

\begin{figure}[!htbp]
  \centering
  \includegraphics[width=0.8\linewidth]{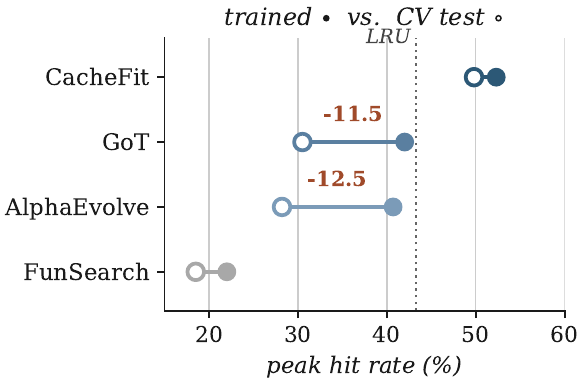}
  \caption{Prior methods produce policies near the published canon. Perf-SHiP++ sits well outside the imitation
  zone and loses only $2.5$ points. The two effects are connected. }
  \label{fig:teaser}
\end{figure}

In the second Cache Replacement Championship, MPPPB ranks second on single core %
by less than 0.01\% speedup difference from the winner. Fit-MPPPB achieves 1.1\% over MPPPB while cutting metadata by 60\% for the same configuration. More interestingly this result is achieved from only 12 iterations of \sys{} while championship winners are often the results of months of human tuning; MPPPB itself was tuned extensively with hill-climbing algorithm \cite{mpppb} 
over 10 years of single CPU computing hours. Our results in \S\ref{sec:controlled} (Figure~\ref{fig:teaser}) show the final policy could generalize to unseen workloads as well. These together suggest \sys{} has significant potential to be adapted and assist humans.   

The latency axis did not move.
Victim-path latency changes from $284$ to $283$\,ps. The array read follows the
smaller tables ($87\to86$\,ps), while the reported logic term remains
$197$\,ps. Both headers time the same 
two table reads, one addition, and the
same 64-bit trace-construction network. 

MPPPB's cost, like \sip{}'s, was a sizing and placement
choice rather than a tax paid for prediction quality---and the two baselines
share very little mechanism, which is what makes \sys{} a general tool for \textit{fitting} replacement policies within constraints.

\medskip\subsection{Prefetching}
\label{sec:prefetch}

The main study uses
ChampSim CRC2~\cite{champsim}, the frozen championship build, whose prefetchers
are compiled in and cannot be swapped; this ablation therefore runs on upstream
ChampSim~\cite{champsim}, whose modular prefetcher interface allows the
substitution. For this, Berti~\cite{berti2022} replaces next-line at the L1D;
next-line moves down to the L2 in place of PC-based stride; and the Entangling
Instruction Prefetcher~\cite{ros2021entangling}---an instruction-side
prefetcher with no counterpart in the baseline configuration---is added at the
L1I. The same change was applied to both policies, with the LLC replacement
policy and every other simulation parameter held fixed
(Table~\ref{tab:simconfig-app}).

The advantage does not erode. Both policies gain $6.2\%$ IPC from the stronger
prefetchers---\mpppb{} $1.034 \rightarrow 1.099$, Fit-MPPPB $1.034
\rightarrow 1.098$---and both gain about a quarter of a point of LLC hit rate
(Table~\ref{tab:prefetch-ablation}). 
Fit-MPPPB's storage, area,
and power advantage over \mpppb{} therefore persists
independent of the prefetching configuration, not an artifact of a weak one.

\medskip\begin{table}[!htbp]
\renewcommand{\arraystretch}{1.05}
\caption{Prefetcher Impact}
\label{tab:prefetch-ablation}
\centering
\small
\setlength{\tabcolsep}{4pt}
\begin{tabular}{@{}llrr@{}}
\toprule
Policy & Prefetchers & IPC & LLC HR \\
       &             & (harm.) & (arith.) \\
\midrule
\mpppb{}    & NL@L1D, Stride@L2          & 1.034 & 20.85\% \\
\mpppb{}    & EIP@L1I, Berti@L1D, NL@L2  & 1.099 & 21.10\% \\
\addlinespace[2pt]
Fit-MPPPB   & NL@L1D, Stride@L2          & 1.034 & 20.80\% \\
Fit-MPPPB   & EIP@L1I, Berti@L1D, NL@L2  & 1.098 & 21.08\% \\
\bottomrule
\end{tabular}
\par\vspace{2pt}{\footnotesize\raggedright NL $=$ next-line. Run at 100\,M warmup $+$ 100\,M instructions rather than the 100\,M $+$ 1\,B used elsewhere; preliminary pending a full-length re-run.\par}
\end{table}

\paragraph{Fit-SHiP++ removes replicated prediction state, not the replacement foundation}
 \sip{} and
Fit-SHiP++ retain the same \(65{,}536\)-bit RRPV array. That array alone is
\(91.4\%\) of Fit-SHiP++'s \(71{,}712\)-bit footprint, so the search did not
obtain its saving by weakening the recency state that identifies the victim.

\sip{} spends \(98{,}304\)
bits on per-line signatures, reuse feedback, and the SHCT. Fit-SHiP++ replaces
them with a \(4{,}096\)-bit global PC predictor, a \(2{,}048\)-bit per-set
selector, and a 32-bit random-state register (for BRRIP). Its predictive state therefore
falls from \(98{,}304\) to \(6{,}176\) bits, a \(93.7\%\) reduction; total
metadata falls by \(56.2\%\), and static power falls by \(60.9\%\).

The result isolates the design move. Fit-SHiP++ preserves the shared RRPV
mechanism but moves prediction from state replicated with cache lines to state
held globally or per set.

\medskip\paragraph{One per-line signature can dominate the entire policy}
The strongest comparison comes from three variants of the same mechanism
family. Fit-SHiP++ consumes \(0.0447\)\,mW. LiteSHIP-SRRIP, which attaches a
12-bit PC signature to every line, consumes \(0.471\)\,mW. Their static power
differs by \(10.5\times\), while their IPC differs by only \(1.2\%\).
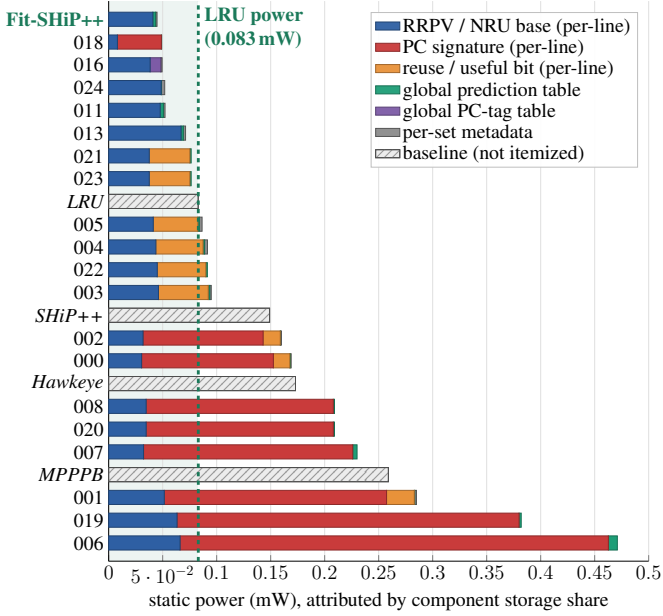
\begin{figure}[t]
\centering
\definecolor{rrpvblue}{HTML}{2E5FA3}
\definecolor{pcsigred}{HTML}{C93B3B}
\definecolor{reuseorange}{HTML}{E8923A}
\definecolor{gpredgreen}{HTML}{1E9E77}
\definecolor{gtagpurple}{HTML}{7D5BA6}
\definecolor{persetgray}{HTML}{8C8C8C}
\definecolor{lruteal}{HTML}{1B7A5A}

\pgfplotstableread[col sep=comma, string type]{
policy,rrpv,pcsig,reuse,gpred,gtag,perset,base,row
Fit-SHiP++,0.0409,0,0,0.0026,0,0.0013,0,0
018,0.0081,0.0407,0,0.0003,0,0,0,1
016,0.0384,0,0,0,0.0099,0.0012,0,2
024,0.0486,0,0,0.0003,0,0.0030,0,3
011,0.0477,0,0,0.0030,0,0.0015,0,4
013,0.0670,0,0,0.0021,0,0.0021,0,5
021,0.0376,0,0.0376,0.0012,0,0,0,6
023,0.0376,0,0.0376,0.0012,0,0,0,7
LRU,0,0,0,0,0,0,0.0829,8
005,0.0413,0,0.0413,0.0013,0,0.0026,0,9
004,0.0437,0,0.0437,0.0014,0,0.0027,0,10
022,0.0450,0,0.0450,0.0014,0,0,0,11
003,0.0461,0,0.0461,0.0014,0,0.0014,0,12
SHiP++,0,0,0,0,0,0,0.1490,13
002,0.0318,0.1112,0.0159,0.0001,0,0.0010,0,14
000,0.0305,0.1220,0.0153,0.0002,0,0.0010,0,15
Hawkeye,0,0,0,0,0,0,0.1730,16
008,0.0347,0.1733,0,0.0011,0,0,0,17
020,0.0347,0.1733,0,0.0011,0,0,0,18
007,0.0323,0.1937,0,0.0040,0,0,0,19
MPPPB,0,0,0,0,0,0,0.2590,20
001,0.0515,0.2058,0.0257,0.0004,0,0.0016,0,21
019,0.0633,0.3167,0,0.0020,0,0,0,22
006,0.0661,0.3966,0,0.0083,0,0,0,23
}\powertable

\resizebox{\columnwidth}{!}{%
\begin{tikzpicture}
\begin{axis}[
    xbar stacked,
    width=15cm, height=15.5cm,
    bar width=10pt,
    xmin=0, xmax=0.5,
    ymin=-0.8, ymax=23.8,        %
    y dir=reverse,               %
    ytick={0,1,...,23},
    yticklabel style={font=\Large},
    xticklabel style={font=\Large},
    yticklabels={%
        {\bfseries\color{lruteal}Fit-SHiP++},018,016,024,011,013,021,023,
        {\itshape LRU},005,004,022,003,
        {\itshape SHiP++},002,000,
        {\itshape Hawkeye},008,020,007,
        {\itshape MPPPB},001,019,006},
    xlabel={static power (mW), attributed by component storage share},
    xlabel style={font=\Large},
    axis x line*=bottom, axis y line*=left,
    xmajorgrids, grid style={gray!25},
    set layers,
    legend style={
        at={(0.98,0.98)}, anchor=north east,
        font=\Large, cells={anchor=west},
        draw=gray!50, fill=white, fill opacity=0.95, text opacity=1,
    },
    legend cell align=left,
    area legend,
]

\addplot[fill=rrpvblue,    draw=rrpvblue!70!black]
    table [x=rrpv,   y=row] {\powertable};
\addplot[fill=pcsigred,    draw=pcsigred!70!black]
    table [x=pcsig,  y=row] {\powertable};
\addplot[fill=reuseorange, draw=reuseorange!70!black]
    table [x=reuse,  y=row] {\powertable};
\addplot[fill=gpredgreen,  draw=gpredgreen!70!black]
    table [x=gpred,  y=row] {\powertable};
\addplot[fill=gtagpurple,  draw=gtagpurple!70!black]
    table [x=gtag,   y=row] {\powertable};
\addplot[fill=persetgray,  draw=persetgray!60!black]
    table [x=perset, y=row] {\powertable};
\addplot[
    pattern=north east lines, pattern color=gray!80,
    draw=gray!60!black,
    preaction={fill=gray!15}]
    table [x=base,   y=row] {\powertable};

\legend{RRPV / NRU base (per-line),
        PC signature (per-line),
        reuse / useful bit (per-line),
        global prediction table,
        global PC-tag table,
        per-set metadata,
        baseline (not itemized)}

\pgfplotsextra{%
\begin{pgfonlayer}{axis background}
\fill[lruteal!8]
    (axis cs:0,-0.8) rectangle (axis cs:0.083,23.8);
\end{pgfonlayer}
\begin{pgfonlayer}{axis foreground}
\draw[lruteal, dashed, line width=2pt]
    (axis cs:0.083,23.8) -- (axis cs:0.083,-0.8)
    node[anchor=north west, font=\Large\bfseries, align=left]
    {LRU power\\(0.083\,mW)};
\end{pgfonlayer}
}
\end{axis}
\end{tikzpicture}
}%
\caption{Component-attributed static power for all $24$ policies, sorted by power; per-line PC-signature arrays (red) dominate cost.}
\label{fig:component-tradeoff}
\end{figure}

The signature array in LiteSHIP-SRRIP consumes \(0.397\)\,mW by itself. That
single array costs more than \sip{}'s complete replacement-policy budget and
nearly \(9\times\) Fit-SHiP++'s entire budget. Both policies use PC information.
The expensive choice is not whether to use the PC; it is whether PC identity
is carried with every cache line.

The broader component breakdown in
Figure~\ref{fig:component-tradeoff} supports the same mechanism. Normalized to
LRU, global and per-set structures account for \(0.02\times\) its power,
narrow per-line fields account for \(0.41\)--\(0.53\times\), and a per-line PC
signature accounts for \(2.32\times\). Across policies that contain them, the
attributed power averages \(0.193\)\,mW for a signature array and
\(0.002\)\,mW for a global prediction table. Changing component families moves
median IPC by only \(0.0007\), yet moves IPC/mW by as much as \(4.7\).

\medskip\paragraph{Fit-MPPPB exposes a different form of slack}
\sip{} makes an insertion decision from one indexed observation:
\[
    \mathrm{SHCT}[\mathrm{Hash}(PC)].
\]
Shrinking that table causes previously distinct PCs to share the same counter.
Their updates then overwrite the only observation used by the decision. The
smaller table does not merely reduce precision; it changes which histories the
policy can distinguish.

MPPPB instead combines \(K\) independently indexed observations:
\[
    y =
    \sum_{i=1}^{K}
    W_i\!\left[H_i(PC,\mathrm{features})\right].
\]
A collision in one table need not occur in the others. Reducing \(K\) therefore
removes some evidence without replacing all evidence for the access.

\sys{} exploits that structure through three changes:
$
K:16\rightarrow8,\qquad
\text{weight width}:6\rightarrow5\ \text{bits},\qquad \
\text{sampler}:80{\times}18\rightarrow64{\times}16.
$
These changes reduce the number and precision of the votes but preserve the
same hash--accumulate--threshold decision rule.

The measured behavior follows that mechanism. Fit-MPPPB matches or exceeds
baseline MPPPB's hit rate on 19 of 20 workloads and its IPC on 16 of 20. Every
IPC regression is below \(2.3\%\), and no workload loses both IPC and hit
rate. It also improves IPC per kilobyte on every workload: by \(2.6\times\) under the mean of per-workload ratios and by \(2.5\times\) under the ratio of
aggregate IPC to aggregate storage.

\medskip\subsection{Comparing with other AI frameworks}
The generalization gap separates the frameworks more sharply than
trained performance does. GoT and AlphaEvolve reach competitive peak
hit rates when trained ($42.0\%$, $40.7\%$) but lose $11.5$ and
$12.5$ points on held-out folds, falling below untrained LRU
($43.3\%$): they discover workload features, not a policy. FunSearch
overfits less but is never competitive. \sys{} alone exceeds LRU in
both regimes, giving up just $2.5$ points under cross-validation ---
it recovers replacement principles that transfer, not workload
signatures (Figure~\ref{fig:teaser}). The chief enabler is
crossover between embedding-distant parents
(Features~2--3): removing it collapses the
proposals back onto the canon, into the imitation zone the others
occupy.

\medskip\section{Example of Auditability and User Steering}
\label{sec:controlled}

Application of AI in computer architecture has historically shown useful empirical results but has been rarely deployed or widely applied in industry for design, due to trust challenges AI has always had: overfitting, inability to interpret its output, and, more importantly, the inability to audit, fix, and align its functionality. \sys{} is, to our knowledge, the first application of LLMs (and more generally AI) in the field of design that does not rely solely on the LLM output and allows targeted expert knowledge/constraints injected into the learning and system explores the space, edit, augment and restart the search from the point of injection instead of retraining or starting the search over.

 An LLM that generates hardware has no ground truth about
microarchitecture---it has patterns learned from text. Some of
those patterns are correct. Some are not. The only way to tell the
difference is to let a human expert read what the model believes
and judge it. 

 Because every principle is a readable sentence in a file, an
  expert can inspect the agent's reasoning, spot the error, and
  trace exactly which entries caused which behavior. In a system
 with conventional NN or LLM frameworks without lessons and principles (such as FunSearch, AlphaEvolve, ArchAgent, Agentic Architect), an error or hallucination or an undesired property of the output replacement policy (not necessarily an error) would persist undetected. 

  In this section we provide an example to clearly show how \sys{} despite being an automated loop, enables an easy and effective method for user to audit and steer the LLM behavior to the desired constraints or special behavior enabling a rich framework for user to explore the domain with \textit{cooperation with \sys{} } instead of complete fully-trusted  design. 

\begin{table}[t]\centering\small
\caption{Four controlled \sys{} runs. User fixes after audits reruns. Controlled study on  V/U $=$ victim- vs.\ update-path memory entries.}
\label{tab:runs}
\setlength{\tabcolsep}{3.5pt}
\begin{tabular}{c l c c c c}
\toprule
\textbf{Run} & \textbf{Config} & \textbf{Iter} & \textbf{Final (ps)}
  & \textbf{IPC} & \textbf{V/U} \\
\midrule
1 & No cap + small-cmp hint      & 25 & 509 & 0.6142 & 1:2 \\
2 & Soft lat500 (score only)     & 25 & 509 & 0.6136 & 1:5 \\
3 & $\leq$500\ps{} + small-cmp   & 20 & 245 & 0.6141 & 7:1 \\
4 & Strict $\leq$500\ps{} prompt & 20 & 267 & 0.6145 & 7:1 \\
\bottomrule
\end{tabular}
\end{table}

 Table~\ref{tab:runs} lists four controlled \sys{} runs. 
The configuration is identical across runs—same baseline policy, same workloads, same scoring function—except for the latency text in the prompt and the presence or absence of an explicit victim-path cap. The only thing that changes is what the designer tells \sys{} about the allowable victim-path delay.
In Run~1 and Run~2 (no cap or soft cap), the memory fills predominantly with update-path entries (U) describing richer predictors, PC features, and learning heuristics, while victim-path (V) discipline remains sparse. The loop converges to designs with strong IPC but long victim-path latency (509\ps{}), reproducing the community’s bias toward on-path aging and per-line state. In Run~3 and Run~4 (strict $\leq$500\ps{} constraint with stage-level attribution), the balance flips: the memory now accumulates V entries such as “defer nothing to the victim path” and “no fold-min, saturating add, or aging cascade on the critical path,” and the resulting policies shift prediction and learning off the victim path while keeping victim selection as a narrow comparison. The final latencies drop to 245–267\ps{} with essentially unchanged IPC.

 After 25 iterations on the latency-constrained SPEC\,2017 run,
\sys{} has written the following entries into its SQLite memory
file. Any user can open this file and read them:

\begin{enumerate}[leftmargin=1.2em,itemsep=3pt]
\item \textbf{PRINCIPLES:} \emph{``Defer nothing to the victim
  path; place prediction intelligence on the update path and
  consume a precomputed result on the critical path.''}

\item \textbf{DONTS:} \emph{``No fold-min, sat-add, or aging
  cascade on the victim path---these are the 286\,ps of
  \shippp{} post-SRAM logic that block the 500\,ps latency
  constraint.''}

\item \textbf{PRINCIPLES:} \emph{``Parallel SRAM reads of rrpv /
  reuse / hot fields add zero latency over a single read.''}

\item \textbf{DONTS:} \emph{``No on-path SHCT lookup---move
  signature-history queries to the update path.''}

\item \textbf{KNOBS:} \emph{``Two bits of RRPV per line is
  sufficient; three or more adds area with no IPC gain.''}
\end{enumerate}

 Entries~1 and~2 dominate the agent's behaviour. They
tell it to keep the victim path as simple as possible and move all
prediction logic to the update path. The agent obeys: it generates
policies with short, fast victim paths (as low as $245$\,ps) and
avoids any complex logic on the eviction path.

 This V/U ratio flip is an audit trail and a steering mechanism. First, the designer can open the memory file and see, line by line, which principles the system has inferred from prior runs: whether it is spending  budget on predictor-side features (U) or on victim-path constraints (V). Second, the designer can intervene. A two-row INSERT into memory immediately redirects the trajectory of a run from a long victim path to a short one; a symmetric DELETE regrows the expensive chain. In other words, when the user audits the memory and changes it, \sys{} obeys.

The expert deletes entries~1, 2, and~4, and replaces
them with deployment-aware principles. For example, the expert
can elevate \emph{cache coherence}, \emph{prefetch interaction},
and \emph{off-chip traffic} as primary design concerns. Or the expert can replace the PC-dependent entry with
a principle that acknowledges the routing cost: \emph{``Do not depend on the program counter at the LLC.
  The PC is expensive to route from the front end, and prefetch
  requests do not carry one. Use access-pattern signatures,
  memory-region metadata, or local reuse counters instead.''}~\cite{kim2017killpc}. 

Without this editable memory, an LLM-driven search would remain a black box: it could wander into highly specialized corners of the design space—e.g., elaborate SRRIP variants with large per-line metadata—that look attractive under IPC-only scoring but are misaligned with hardware feasibility and the human architect’s intent. By exposing the system’s principles as text and making them the substrate that actually shapes future proposals, \sys{} turns AI from an uncontrollable generator into a co-designer that a human can guide, correct, and specialize. This is the key step that converts “AI for architecture” from a one-off optimizer into a human–AI assistant for ongoing microarchitectural design.

\medskip\section{Conclusion}
\label{sec:conclusion}

Across published LLC replacement policies, logic depth varies by $6.1\times$
and power per kilobyte by $1.7\times$: metadata capacity alone cannot
predict relative implementation cost and misorders the policies. We showed that even for a given \textit{known} policy, redesigning it to reduce the cost manually or with known LLM-assist tools is non-trivial and often fails; producing infeasible and largely complex policies. 
We thus focus this first study on improving only known replacement policy methods, e.g. MPPPB-type or SHiP++-type, and exclude other novel policies beyond known method variations exploration from this study.

\sys{} exposes efficiency-oriented design choices through its learned principles as it explores the space and
turns them into \textit{fit} versions of each policy---among them Fit-MPPPB, which cuts storage of the given policy (the seed) by
$60.3\%$, power by $56.3\%$, and transistors by $58.9\%$, while improving
IPC on 14 of 20 SPEC~CPU2017 workloads and hit rate on 18 of 20; making cost efficiency a direct objective in performance-seeking design-space explorations.

\bibliographystyle{IEEEtranS}
\bibliography{refs}

\end{document}